\documentclass[twocolumn,letterpaper]{IEEEAerospaceCLS}  % only supports two-column, letterpaper format

\usepackage[]{graphicx}    % We use this package in this document
\usepackage{multirow}
\usepackage[hyphens,spaces,obeyspaces]{url}
\usepackage[colorlinks,allcolors=blue]{hyperref}
\usepackage[table,xcdraw]{xcolor}
\usepackage{tikz}
\newcommand{\ignore}[1]{}  % {} empty inside = %% comment
\usepackage[utf8]{inputenc} %to use Hawaiian diacritical marks  
\usepackage{newunicodechar}     % """"
\DeclareRobustCommand{\okina}{%
  \raisebox{\dimexpr\fontcharht\font`A-\height}{%
    \scalebox{0.8}{`}%
  }%
}
\newunicodechar{ʻ}{\okina}

\begin{document}
\title{Jitter Characterization of the HyTI Satellite}

\author{
Chase Urasaki\\ 
Department of ECE \\
University of Hawai\okina i\\
Honolulu, HI 96822\\
chasemu@hawaii.edu
\and 
Frances Zhu\\
Hawai\okina i Institute for Geophysics and Planetology\\
University of Hawai'i \\
Honolulu, HI 96822\\
zhuf@hawaii.edu
\and
Michael Bottom\\ 
Institute for Astronomy\\
University of Hawai'i \\
Honolulu, HI 96822\\
mbottom@hawaii.edu
\and 
Miguel Nunes \\ 
Hawai\okina i Space Flight Laboratory\\
University of Hawai\okina i\\
Honolulu, HI 96822\\
manunes@hawaii.edu
\and 
Aidan Walk \\ 
Subaru Telescope \\
National Astronomical Observatory of Japan \\ 
Hilo, HI 96720 \\
walk@naoj.org\\
%%%% IMPORTANT: Use the correct copyright information--IEEE, Crown, or U.S. government. %%%%%
\thanks{\footnotesize 979-8-3503-0462-6/24/$\$31.00$ \copyright2024 IEEE}              % This creates the copyright info that is the correct 2024 data.
%\thanks{{U.S. Government work not protected by U.S. copyright}}         % Use this copyright notice only if you are employed by the U.S. Government.
%\thanks{{979-8-3503-0462-6/24/$\$31.00$ \copyright2024 Crown}}          % Use this copyright notice only if you are employed by a crown government (e.g., Canada, UK, Australia).
%\thanks{{979-8-3503-0462-6/24/$\$31.00$ \copyright2024 European Union}}    % Use this copyright notice is you are employed by the European Union.
}

\maketitle

\thispagestyle{plain}
\pagestyle{plain}

\maketitle

\thispagestyle{plain}
\pagestyle{plain}

\begin{abstract}
The Hyperspectral Thermal Imager (HyTI) is a technology demonstration mission that will obtain high spatial, spectral, and temporal resolution long-wave infrared images of Earth’s surface from a 6U cubesat. HyTI science requires that the pointing accuracy of the optical axis shall not exceed 0.014 mrad (approximately 2.89 arcseconds) over the 0.5 ms integration time due to these effects (known as jitter). Two sources of vibration are a cryocooler that is added to maintain the detector at 68 K and three orthogonally placed reaction wheels that are a part of the attitude control system. Both of these parts will introduce vibrations that get propagated through to the satellite structure while imaging. Typical methods of characterizing and measuring jitter involve complex finite element methods, computationally expensive modeling, expensive equipment and specialized laboratory setups. In this paper, we describe a novel method of characterizing jitter for small satellite systems that is low-cost, simple, and minimally modifies the subject’s mass distribution. The metrology instrument is comprised of a laser source, a small mirror that is mounted via a 3-D printed clamp to a jig, and a lateral effect position-sensing detector.  The position-sensing detector samples 1000 Hz and can measure displacements as little as 0.15" at distances of one meter. This paper provides an experimental procedure that incrementally analyzes vibratory sources to establish causal relationships between sources and the vibratory modes they create. We demonstrate the capabilities of this metrology system and testing procedure on HyTI, using the advanced Attitude Determination, Control, and Sensing (ADCS) Test Facility in the Hawaii Space Flight Lab’s clean room. Results include power spectral density plots that show fundamental and higher-order vibratory modal frequencies in HyTI with a precision of better than one arcsecond measured at distances of approximately one meter. The metrology instrument and procedure can attribute correlation and possibly causation of these modal frequencies to vibratory sources. Results from metrology show that jitter from reaction wheels meets HyTI system requirements within 3$\sigma$. 
\end{abstract} 

\tableofcontents

%%%%%%%%%%%%%%%%%%%%%%%%%%%%%%%%%%%%%%
\section{Introduction}
%%%%%%%%%%%%%%%%%%%%%%%%%%%%%%%%%%%%%%
In recent years, the compact size, cost-effectiveness, and improved technological abilities of cube satellites (cubesats) have increased opportunities for universities and organizations to conduct experiments and develop payloads that were not feasible before. The Hyperspectral Thermal Imager (HyTI) is one such mission and will serve as a technology demonstration to obtain high spatial and spectral resolution long-wave infrared images from a 6U cubesat \cite{HyTI_2019} \cite{HyTI_2022}. 

The Earth Science Decadal Survey identifies high spatial and either multi- or hyper-spectral thermal infrared data as a measurement approach for hydrology, ecosystems, weather, climate, and the solid Earth \cite{CAWSENICHOLSON2021112349}. For example, scientists who wish to map the chemistry of rocks and minerals on the Earth’s surface, study the composition of volcanic gas and ash plumes, or quantify soil moisture content and evapotranspiration rates in the long-wave infrared are limited to approximately 60 - 120 meters of ground sampling in no more than five spectral bands between 8.125 $\mu$m and 11.65 $\mu$m from missions such as Terra ASTER \cite{HyTI_2019} \cite{HyTI_2022}. HyTI will give scientists greater access to these regions of the spectrum by providing 25 spectral bands between 8$\mu$m and 10.7 $\mu$m to study more of Earth’s processes in further detail. 

To study these effects, HyTI's payload is a Fourier transform spectrometer that features a novel no-moving-parts imager that consists of two pieces of germanium separated by an air gap that forms a Fabry-Perot interferometer. This interferometer then images on JPL’s Barrier InfraRed Detector Focal Plane Array (BIRD FPA). BIRD FPAs are antimony compound-based detectors and feature a high signal-to-noise ratio, pixel uniformity, and temporal stability at relatively high operating temperatures for an IR detector \cite{HyTI_2019}, which offers a new way to realize low-cost and high-performance missions. The forward motion of the satellite along the “in-track” direction allows ground targets to be imaged at different optical path lengths to generate an interference pattern. The resulting images are co-registered, and a spectro-radiometrically calibrated cube is generated using standard Fourier transform techniques. 

%%%%%%%%%%%%%%%%%%%%%%%%%%%%%%%%%%%%%%%%%
%Shcematic of how hyti works - ask Miguel. is this a pushbroom optical setup>?

An SF070 single-stage Stirling cryocooler from AIM Infrarot-Module is necessary to maintain the detector at 68 K to achieve acceptable dark current levels \cite{HyTI_2019} \cite{HyTI_2022}. To meet requirements for electromagnetic interference, a Creare Micro-cryocooler Control Electronics for Tactical Space and a Creare/West Coast Solutions Active Ripple Filter were also added.

Ultimately, the quality and reliability of the spectra obtained from the payload will depend on the stability of the satellite platform. Thus, one of the important science and technology requirements for the HyTI mission is that the pixel blur shall not exceed 10\% of the instantaneous field of view (IFOV) over 0.5 ms - the nominal integration time of the payload. This corresponds to a maximum acceptable pointing error of 0.014 mrad (approximately 2.89 arcseconds). This pixel blur is also known as jitter. Jitter comes from micro-vibrations that originate from the movement or vibration of components on or within the satellite that transfer through the satellite’s structure and propagate into the payload. We expect that the main contributors of jitter for HyTI are the unbalanced reciprocating pistons in the cryocooler and the imbalance in the reaction wheel motors as they are the only moving components in the spacecraft.

\begin{figure}[hbt!]
    \centering
    \includegraphics[width=\linewidth]{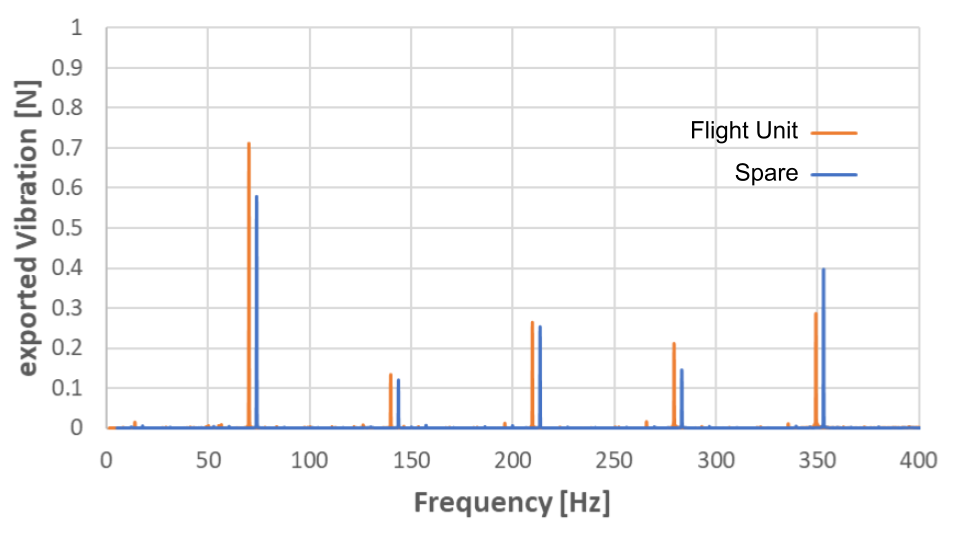}
    \caption{ Exported vibration (in Newtons) in the compressor drive axis as a function of frequency measured from the two cryocoolers that were selected. Both coolers were operated at 65 Hz, with the data from the spare cryocooler shifted by +4 Hz for clarity.}
    \label{fig:exported_vibe}
\end{figure}

\begin{figure}[hbt!]
\caption{CAD model of HyTI featuring the SF070 cryocooler placed in the center of the satellite to passively mitigate as much off-axis vibration as possible.}
    \centering
    \includegraphics[width =\linewidth]{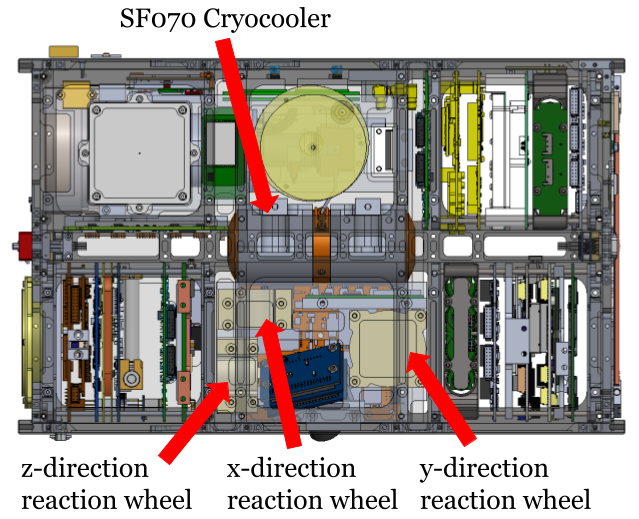}
    
    \label{fig:HyTI_cad}
\end{figure}

Two of the best-performing cryocoolers that met instrument requirements with adequately low vibrations were selected from serial production and were delivered by AIM. Figure \ref{fig:exported_vibe} shows the exported vibrations for the flight unit cryocooler and the spare. The cryocooler was placed near the center of the HyTI to passively eliminate as much of these exported vibrations in operation as possible, as shown in Figure \ref{fig:HyTI_cad}.

The CubeSpace reaction wheels provide attitude control in three independent axes and are labeled in Figure \ref{fig:HyTI_cad}. This allows the Attitude Determination and Control Subsystem (ADCS) to point with a precision of less than 0.1 degrees in conjunction with the CubeStar star tracker. The primary source of disturbances by reaction wheels can be attributed to mass imbalances, bearing imperfections, and motor properties \cite{MASTERSON2002575}. Rotating reaction wheels generate a variety of sub-harmonics and higher-order harmonics from bearing interactions and differ from disturbances from cryocoolers \cite{Dennehy_Alvarez-Salazar_2018}.
\textbf{The inevitable operation of the cryocooler and reaction wheels during data collection means that the jitter caused by these sources is a fundamental limiting factor for the pointing and imaging performance of not only HyTI, but any other future mission that incorporates these solutions.} 

This paper introduces a novel method to characterize jitter given a fully integrated system without the need for a finite element model that can be measured on-ground. In Section \ref{sec:background}, current methods of measuring jitter are briefly described. In \ref{sec:methodology}, the metrology system, testing facility and environment,  data collection process, and equations used to analyze the data are described. Section \ref{sec:Results} presents the results of initial testing and current limitations. Section \ref{sec:conclusions} discusses conclusions and future work. 

%%%%%%%%%%%%%%%%%%%%%%
\section{Background} \label{sec:background}
%%%%%%%%%%%%%%%%%%%%%%%
There are a number of ways that jitter is currently measured and characterized on the ground and in space. In disturbance modeling, component-level testing is completed to inform and validate disturbance models. Detailed and robust tests are conducted for the various components of the satellite on the ground. In an experimental study of a 0.1 arcsec space pointing instrument, a three-axis reaction wheel was attached to the imaging payload which was suspended using a quasi-zero stiffness device \cite{LI2023191}. The instrument imaged a point source generated on a vibrationally-isolated platform and the movement of that point source is analyzed in the frequency domain as a function of reaction wheel speeds and combinations. While they were able to identify the jitter response of the imager and the sub-harmonics generated as a function of reaction wheel properties, the system-level reactions are not explored. Moreover, Masterson et al. show that empirical models for estimating reaction wheel vibration data under-predict disturbances, likely due to the fact that the model does not account for amplifications caused by interactions between harmonics and the structural modal frequencies of the wheel \cite{MASTERSON2002575}.

Measurements of force directly by piezoelectric sensors are a popular solution due to their high measurement resolution and signal-to-noise \cite{piezo_transducers}. They can be used to characterize individual components, such as in the ESA Reaction Wheel Characterization Facility \cite{decobert_2012} or entirely integrated systems. Carpenter, Martin, and Hinkle describe a 6 Degree-of-Freedom measurement platform that incorporates six force transducers that are mapped to force and moment measurements at a rate of 10 kHz. Once data was gathered, the spacecraft was modeled by a transfer function that mapped the forces and torques measured to the resulting angular displacement at the payload \cite{carpenter_6dofjitter}. However, these results are fundamentally dependent on an accurate finite element model. 

On-ground jitter analysis of these integrated systems relies on the use of a finite element model that is representative of the entire structural dynamics. This is true for both cubesats and flagship observatories. The challenge is then to accurately capture and represent the dynamics of a structure with hundreds to thousands of closely spaced lightly-damped modes of vibration with predication accuracy decreasing as frequency increases \cite{Dennehy_Alvarez-Salazar_2018}. Loss of accuracy can also occur from the quality and realism made in the model for sub-assembly and sub-component interfaces and interconnection between boundaries \cite{Dennehy_Alvarez-Salazar_2018}. When working with these finite element models, they are typically a very high-order matrix of the system that is difficult to gain informative physical insight from directly. 

The true test of the accuracy of a model is the comparison with the information obtained in orbit. There are three approaches to obtaining jitter measurements in orbit. The first is to use a high-performance attitude sensor that can deliver a high sampling rate and high accuracy \cite{sabelhaus2001orbit}. The second method is to measure the movement of image points through an additional image sensor on the focal plane as seen in obtained overlapping images by using an auxiliary high frame rate imaging sensor on the focal plane and an optical correlator \cite{Janschek_2005}. The third method utilizes parallax observations between two slightly offset sensors in a pushbroom payload that can acquire overlapping images at slightly different times. This method does not require any additional high-performance sensors like the previous two methods discussed and as a result, is a popular method for estimating, detecting, and validating jitter for payloads with CCDs \cite{Liu:19}. In-orbit techniques have the advantage of representing the true jitter that the satellite experiences and provide the most accurate results as they pertain to mission operations while not requiring the need for a finite element model. On the other hand, it cannot characterize jitter on a component or assembly level and does not allow for the manipulation or alteration of system parts to achieve more optimal jitter characteristics. 

%%%%%%%%%%%%%%%%%%%%%%%%%%%%%%%%%%%%%%%%%%%%%%%%%%
\section{Methodology} \label{sec:methodology}
%%%%%%%%%%%%%%%%%%%%%%%%%%%%%%%%%%%%%%%%%%%%%%%%%%
\subsection{Metrology System} \label{sec:setup}
\begin{figure}[hbt!]
    \centering
    \includegraphics[width = \linewidth]{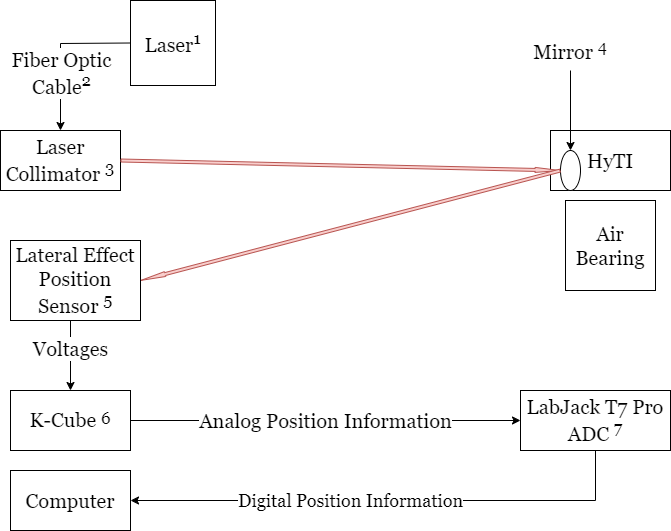}
    \caption{Block diagram schematic of the metrology setup. Superscripts link the component to the detailed description in the text. The full detail of the setup is described in Urasaki et. al (in prep).}
    \label{fig:block_diagram}
\end{figure}

%%%%%%%%%%%%%%%%%%%%%%%%%%%%%%%%%%%%%%%%%
The concept behind this optical, low-cost metrology setup is to measure the jitter of HyTI by using a detector to measure and record small deflections of stationary laser beam off of a mirror that is attached to the satellite. To achieve low cost and replicability, commercial off-the-shelf (COTS) components were used to develop the metrology system, each superscript in the text is linked to a node or connection in Figure \ref{fig:block_diagram}. 

\begin{figure}
    \centering
    \includegraphics[width =\linewidth, height = 3.7in]{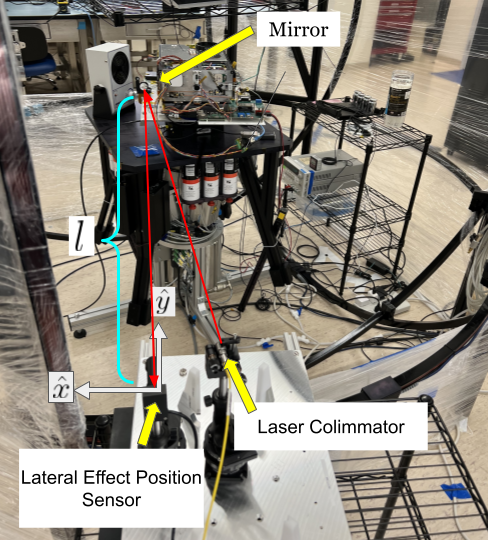}
    \caption{The metrology setup featured with HyTI coupled with support electronics on the air bearing, with $\hat{x}$ indicating the $x$ direction data, $\hat{y}$ indicating the $y$ direction data, and the throw distance $l$.}
    \label{fig:diagram_with_pic}
\end{figure}

The S1FC635PM fiber-coupled laser source$^1$ from Thorlabs is used with a single-mode fiber optic cable $^2$ to feed a laser collimation package$^3$ that is secured to a kinematic v-mount. A standard two-adjuster kinematic mirror mount holds the mirror$^4$ and is rigidly mounted to the supported jig with a 3-D printed clamp.
The Thorlabs Lateral Effect Position Sensor$^5$ measures the displacement of the incident beam relative to the calibrated center of the detector. This sensor was chosen due to its high accuracy and ability to provide positional information of any spot within the detector region, independent of beam shape, size, and power distribution. This obviates the need for optical gain calibration from other wavefront sensors or an expensive, high-speed, pixelated detector. At the maximum incident power level, the sensor has a resolution of 0.75 micrometers limited by electronics.
A Kinesis K-Cube position sensing detector and auto aligner$^6$ was used in open loop mode to measure beam position. The K-Cube generates a left-minus-right (X DIFF) signal, a top-minus-bottom (Y DIFF) signal, and a sum (SUM) signal. The LabJack T7-Pro$^7$ was used to convert the analog signal from the K-Cube into reported $x$ and $y$ values (in the dimension of length) for the center of the incident beam with a Nyquist frequency of 500 Hz (half the 1 kHz sampling rate).
These digital position values are then stored in a text file for analysis in Python. 
To convert the position values into an angular displacement in units of arcseconds, the centroid values reported by the setup were divided out by the distance between the mirror and the lateral effect position sensor, $l$. Figure \ref{fig:diagram_with_pic} shows the setup with HyTI on the air bearing along and the throw distance labeled.  

This setup can provide highly accurate measurements of laser beam deflection. Electronic noise limits the lateral effect position sensor to a resolution of $\delta$= 0.75 $\mu$m in both the x and y directions and given $l = 1~\mathrm{m}$ (the approximate throw distance between trials), this corresponds to a minimum measurable angular displacement of 0.154", shown in Equation \ref{eq:angular_reso}. This system is also scalable to measure jitter simultaneously in three degrees of freedom; one would just need to duplicate the mirror to the position sensor path orthogonally to the existing path.
\begin{equation}
    \label{eq:angular_reso}
     d\theta = \frac{\delta}{l}\frac{206265"}{\mathrm{rad}}
\end{equation}

%%%%%%%%%%%%%%%%%%%%%%%%%%%%%%%%%%%%%%%%
\subsection{Jitter Testbed and Environment} \label{sec:testbed}
%%%%%%%%%%%%%%%%%%%%%%%%%%%%%%%%%%%%%%%%

The metrology system was set up alongside the ADCS test facility in the 10,000-class HSFL Cleanroom at the University of Hawai\okina i at M\=anoa. The test facility, built by Astrofein, can provide highly accurate and complete end-to-end verification of ADCS systems for small satellites using a rotary air-bearing platform (to which HyTI was attached) by Physik Instrumente (PI) and a Helmholtz cage to simulate Earth’s magnetic field in orbit. 
The acceptance test from PI for our specific air bearing reports an axial error motion of 0.036 $\mu$m, a radial error motion of 0.058 $\mu$m, and a tilt error of 0.1 arcsec. Using $\delta_{ax} = 0.036~\mu\mathrm{m}$ in Equation \ref{eq:angular_reso} for an $l = 1~\mathrm{m}$ this corresponds to a d$\theta$ = 0.0074", much less than the minimum angular displacement measurable at this throw distance calculated in the previous section and thus, we do not expect the axial error to contribute any systematic error to our measurements. The radial error does not contribute to measurements made since that motion occurs orthogonally to the sensor and does not result in any displacement of the incident beam in the directions measured. The tilt error, which is the combined effect of the axial and radial errors, is on the order of the fundamental precision of our metrology setup.

%insert discussion about dynamic range?   

\begin{figure}[hbt!]
    \centering
    \includegraphics[width = \linewidth]{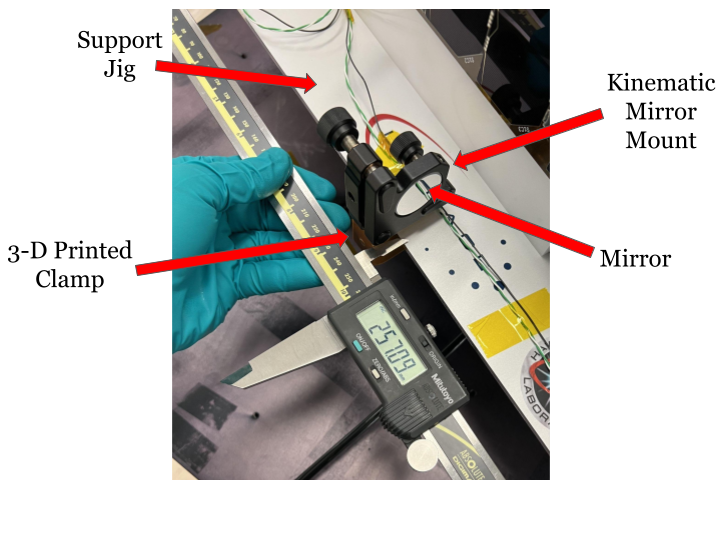}
    \caption{Kinematic mirror mount with the mirror inserted secured to the support jig that rests atop the rotary air bearing. The distance to the center of mass from the front edge of the support jig is 257.09 mm according to CAD modeling. }
    \label{fig:mirror_with_jig}
\end{figure}

A 3-D printed plastic clamp secures the kinematic mount holding the mirror to the 6U support jig. This is shown in Figure \ref{fig:mirror_with_jig}, where calipers were used to place the mirror as accurately as possible to the calculated center of mass on the jig. Errors in this case come from the misalignment between the mirror face and the edge of the calipers which is no more than 0.05 mm. In this case and shown in Figure \ref{fig:diagram_with_pic}, the x-axis of the sensor corresponds to yaw and the y-axis corresponds to pitch. 

\begin{table*}[hbt!]

\caption{Vibratory combinations for which data has been collected. The configurations are ordered such that each configuration has one more vibrational source or component active than the previous one.}
\renewcommand*{\arraystretch}{1.2}
\centering
\begin{tabular}{|c|c|c|c|c|c|c|}
\hline
Configuration &
  Bearing &
  \begin{tabular}[c]{@{}c@{}}Bus\\ Electronics\end{tabular} &
  ADCS Electronics &
  \begin{tabular}[c]{@{}c@{}}ADCS Reaction \\ Wheels (Direction and RPM)\end{tabular} &
  \begin{tabular}[c]{@{}c@{}}Cryocooler and \\ Camera (Payload)\end{tabular} &
  \begin{tabular}[c]{@{}c@{}}Number of \\ Trials\end{tabular} \\ \hline
1 & OFF & ON & OFF & OFF                                                                             & OFF & 5 \\ \hline
2 & OFF & ON & OFF & OFF                                                                             & ON  & 6 \\ \hline
3 & ON  & ON & OFF & OFF                                                                             & OFF & 5 \\ \hline
4 & ON  & ON & ON  & OFF                                                                             & OFF & 3 \\ \hline
5 & ON  & ON & ON  & \begin{tabular}[c]{@{}c@{}}ON \\ (x direction at 3000 RPM)\end{tabular}         & OFF & 3 \\ \hline
6 & ON  & ON & ON  & \begin{tabular}[c]{@{}c@{}}ON \\ (z direction at 3000 RPM)\end{tabular}         & OFF & 3 \\ \hline
7 & ON  & ON & ON  & \begin{tabular}[c]{@{}c@{}}ON\\ (x and z direction \\ at 3000 RPM)\end{tabular} & OFF & 4 \\ \hline
\end{tabular}
\label{tab:sources_tried}
\end{table*}

\begin{figure}[hbt!]
    \centering
    \includegraphics[width=\linewidth]{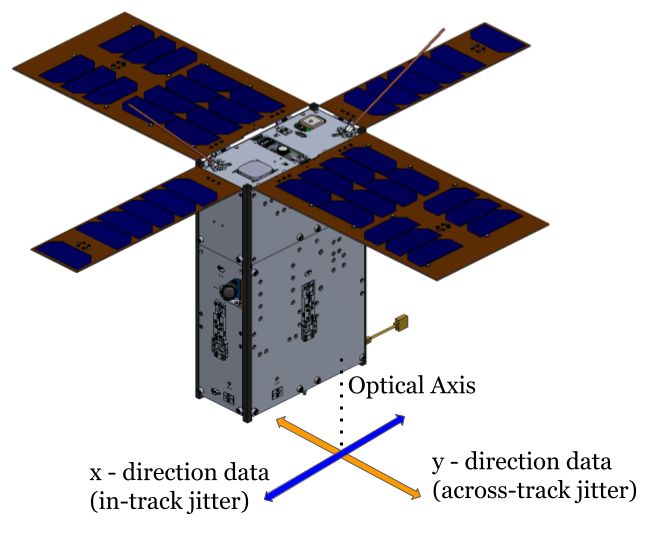}
    \caption{HyTI in its operations configuration where the optical axis forms the origin, the in-track jitter that is recorded by our metrology setup is captured by the x-direction data, and the across-track jitter is captured by the y-direction data.}
    \label{fig:hyti_deployment}
\end{figure}

Damping effects from the 3-D printed plastic clamp are assumed to be negligible and approximated as a rigid body at these frequencies for this work. In orbit and deployment, the jitter in the x- and y-directions will correspond to the jitter in the in-track and across-track dimensions respectively. This is pictured in Figure \ref{fig:hyti_deployment}. 

Air bearing imbalance introduced challenges in the form of undesired oscillatory motion, where the incident beam would occasionally drift off of the sensor area leading to a lack of high signal-to-noise readouts from the K-cube.
%%%%%%%%%%%%%%%%%%%%%%%%%%%%%%%%%%%%%%%%%%%
\subsection{Experiment Design} \label{subsec:Design}
%%%%%%%%%%%%%%%%%%%%%%%%%%%%%%%%%%%%%%%%%
The process used to evaluate the jitter contribution from various sources involves incrementally adding vibrational sources (or active components) and investigating the impacts of the presence (or absence) of certain sources on the detected vibratory modal frequencies through the comparison of their power spectral density (PSD) plots and calculated jitter values. 

The most relevant vibration influences are the ADCS reaction wheels, cryocooler for the payload, and the air bearing (which will isolate the vibratory motion of HyTI from the ADCS test facility when active). Other components include the satellite bus electronics and the ADCS electronics. These electronics have no moving parts so we do not expect to see any change in the PSDs or calculated jitter values when these components are active. Testing with these configurations allows us to gain confidence in the results and indicate the presence of any systematic errors (if any). By incrementally adding these sources, we can compare PSD plots and the calculated jitter values and attribute differences to the addition of a known source.    

Configuration 1 trials were measured with the air bearing off and the satellite bus electronics turned on. All sources of vibration are off and observed vibration responses present in this trial are representative of the ambient environment in the cleanroom. In Configuration 2 the cryocooler and camera were on, actively taking images while the air bearing remained off. Then, any observed changes in the PSD plots between these two configurations would be from the vibrations of the cryocooler. Jitter values found using Equation \ref{eq:jitter} are not representative of the response of the satellite when the air bearing is off since it turns the satellite and bearing structure into a rigid body. We instead these values and compare them between different configurations with the air bearing off to help identify any potential systematic errors with the system. In Configuration 3, the air bearing was on while the bus electronics remained on and all other components were turned off. In a similar manner to the first configuration, this allows for a baseline for subsequent results. In Configuration 4, the ADCS electronics were turned on. Since this is not an active vibration source, any changes between this and the previous configuration will be due to dynamics from the air bearing. In Configuration 5, the x-direction reaction wheel was spun at a constant rate of 3000 rpm. This angular velocity was chosen to try and detect a fundamental modal frequency at 50 Hz and any generated harmonics. In Configuration 6, x-direction reaction wheel was then turned off and the z-direction reaction wheel was spun at 3000 rpm. Finally, in Configuration 7, both reaction wheels were spun at 3000 rpm to investigate their combined effects. This information is summarized in Table \ref{tab:sources_tried} and the following subsections describe how data was collected and analyzed for these configurations.

%%%%%%%%%%%%%%%%%%%%%%%%%%%%%%%%%%%%%%%
\subsection{Data Collection} \label{subsec:data_collection}
One permutation from Table \ref{tab:sources_tried} is selected. Over a three-minute period, the lateral effect position sensor detects and reports the measured power center of the incident laser spot. The data is sampled at a rate of 1kHz. This results in two time series: i) a plot of the laser center as a function of time in the x-direction as shown in Figure \ref{fig:diagram_with_pic}, and ii) a plot of the laser center in the y-direction as a function of time.
The number of trials conducted per configuration is detailed in Table \ref{tab:sources_tried}.

 %%%%%%%%%%%%%%%%%%%%%%%%%%%%%%%%%%%%%%\
 \subsection{Data Analysis}
To investigate the frequency content of the system dynamics, the Fourier transform for each time series was generated. From the Fourier transform, a PSD was calculated using Equation \ref{eq:psd}, where the $G(T,f)$ is the PSD of some zero-mean random signal $x(t)$ and $X(T,f)$ is the Fourier transform of $x(t)$ over the interval $[0, T]$. The resolution of the PSD, given by the inverse of the duration of the time series, is 5.55$\times$10$^{-3}$ Hz. All PSD plots per configuration were averaged to obtain a more representative sample of the data. No binning or interpolation was required between trials since the resolution of the PSD plots was equal. From the PSD plots, our primary interest lies in quantifying the amount of jitter introduced by system dynamics in the form of pointing errors at two frequencies relevant to the HyTI mission: i) the integration time frequency $f_{int}$ = 2000 Hz (0.5 ms), ii) the frame rate frequency $f_{frame}$= 139 Hz (7.2 ms).

Because HyTI's payload is a spatially modulated imaging Fourier Transform spectrometer, the orbital motion of the spacecraft moves scene elements through fixed optical path differences created by the interferometer. Thus, pointing errors in the integration frequency will result in sampling light from a larger area on Earth's surface and the resulting image will appear blurred. Pointing errors in the frame rate frequency will affect the rate at which the interference pattern moves over the ground. This will affect how scene elements move through successive optical path differences and is equivalent to speeding up or slowing down the spacecraft, changing the effective spectral sampling rate. Oversampling at 139 Hz helps to mitigate this problem slightly for HyTI specifically, but the results will be useful to inform future mission design and planning. 

From \cite{Thuc_2018}, the variance is the cumulative power from the frequency of interest to the highest measurable frequency, $f$ shown in Equation \ref{eq:var}. Because variance measurements made in each axis are independent, the aggregate variance is defined by Equation \ref{eqn:var(x+y)}, where  $\sigma^2_x$ is the result of Equation \ref{eq:var} on the in-direction PSD and $\sigma^2_y$ is the result of Equation \ref{eq:var} on the across-track PSD. After the aggregate variance is calculated, we calculate the $n\sigma$ pointing jitter for our frequency of interest by Equation \ref{eq:jitter}. Because the Nyquist frequency for the system is 500 Hz, ($f_{max} = 500$), we will not be able to compute the jitter via Equation \ref{eq:var} for the integration time frequency since $f_{int}$ is less than $f_{max}$.

\begin{figure*}[hbt!]
    \centering
    \includegraphics[width=6in]{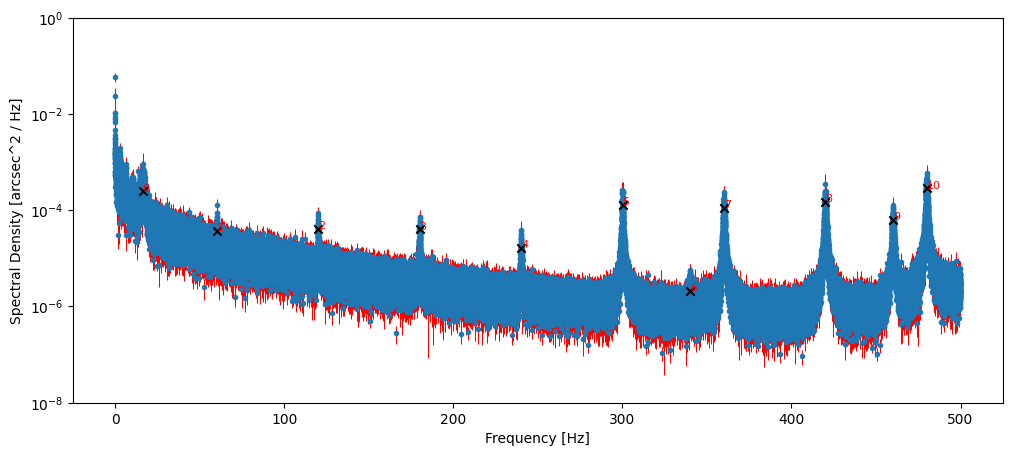}
    \caption{Example averaged PSD plot for Configuration 2 in the in-track direction with peaks identified and marked with black crosses. Error bars for individual points are indicated in red.}
    \label{fig:peaks_id}
\end{figure*}
\begin{equation}
    \label{eq:psd}
    G(T,f) = \frac{1}{T}|X(T,f)|^2
\end{equation}

%\textcolor{red}{double check this section: Mike and I think it was an error in the MIT thesis}.
\begin{equation}
    \label{eq:var}
    \sigma^2(f) = \int_f^{f_{max}}G(T, f')df' 
\end{equation}

\begin{equation}
    \label{eqn:var(x+y)}
    \sigma^2_{(x+y)} = \sigma^2_x + \sigma^2_y
\end{equation}

\begin{equation}
    \label{eq:jitter}
    n\sigma(f) = n\sqrt{\sigma^2_{(x+y)}}
\end{equation}

As an additional objective, we are interested in identifying any fundamental or resonance modal frequencies that originate from individual vibrational sources. An algorithm found peaks and fit Lorentzian profiles to give each modal frequency [Hz], the amplitude [arcsec$^2$/Hz], the full width at half maximum (FWHM) [Hz], and their respective variances. By comparing the differences in the amplitude and the FWHM of common frequencies between configurations, we are able to disambiguate the contributions of different sources at the same frequency. While these values are not useful for calculating the pointing error due to jitter directly, they help accomplish our second objective to better understand system dynamics and inform mission planning and structure and assembly design. The modal frequencies that were identified are compared to those present in the previous configuration. The plot of the peak identification is shown in Figure \ref{fig:peaks_id}.

%%%%%%%%%%%%%%%
\section{Results} \label{sec:Results}
%%%%%%%%%%%%%%%%%
Each subsection indicates the permutation investigated in Table \ref{tab:sources_tried} and presents 3 results: i) the variances as given by Equation \ref{eqn:var(x+y)}, ii) the 1$\sigma$, 2$\sigma$, and 3$\sigma$ values as given by Equation \ref{eq:jitter}, and iii) a table summary of the modal frequencies that were identified in that configuration. The averaged PSD plot for each configuration and their full table of identified modal frequencies including the amplitudes and FWHM are presented in the Appendix. %As a part of the verification plan for the HyTI requirements, simulations calculated the maximum angular displacement around HyTI's across-track dimension as a worst case. These simulations showed that variation of the optical axis was limited to 0.004 arcsec over the integration time and 2683.2 arcsec over the duration of a tile. This value is calculated by taking the simulated angular drift over the tile, 516 arcsec per sec (0.5 mrad/sec), and multiplying it by the tile time. These values met the 2.89 arcsec requirement over the integration time of 0.5 ms and the rate of 1773.88 arcsec per second (8.6 mrad/sec) over the tile time. 
Because the metrology system cannot resolve the integration time jitter directly, we compare the frame rate jitter to the HyTI's requirement for the integration time. Given that $G(T,f)$ is always non-negative, the result of Equation \ref{eq:var} for $f_{frame}$ will always be larger than $f_{int}$ where the integral is valid (i.e. $f_{max}$ is greater than $f$). Thus, since Equation \ref{eq:jitter} is an increasing function, the frame rate jitter bounds the integration time jitter. The code for this analysis is made available \href{https://github.com/chase-urasaki/HyTI_Jitter_Metrology.git}{here} in this GitHub repository. %\textcolor{red}{is there a justification I can use for this? something like integration jitter should always be less than frame rate jitter? My thoughts are that frame rate jitter should bound integration time jitter since the period of the frame rate is larger than the period of the integration time and could result in more angular variation}
 
\subsection{1 - Air Bearing off, Bus Electronics on, ADCS Electronics on, Reaction Wheels off, Cryocooler and Payload off} 
\label{subsec:bearingoff_electronicson_wheelsoff_cryooff}
Configuration 1 inspected the ambient environment of the cleanroom and the air bearing and all active sources were turned off. The variance for the frame rate jitter was 0.002 arcsec$^2$, which yields 1$\sigma$, 2$\sigma$, and 3$\sigma$ values of  0.014 arcsec, 0.028 arcsec, and 0.043 arcsec respectively.
\begin{table}[hbt!]
\centering
\renewcommand*{\arraystretch}{1.1}
\caption{Identified modal frequencies and their sources for Configuration 1. The source of all frequencies present is the ambient cleanroom environment.}
\begin{tabular}{|c|c|c|}
\hline
\begin{tabular}[c]{@{}c@{}}In-Track \\ Frequencies \\ Identified \\ {[}Hz{]}\end{tabular} & \begin{tabular}[c]{@{}c@{}}Across-Track \\ Frequencies\\ Identified\\ {[}Hz{]}\end{tabular} & Source \\ \hline
 & 35 & Environment \\ \hline
 & 70 & Environment \\ \hline
 & 97 & Environment \\ \hline
 & 106 & Environment \\ \hline
 & 120 & Environment \\ \hline
 & 139 & Environment \\ \hline
 & 179 & Environment \\ \hline
 & 211 & Environment \\ \hline
240 & 240 & Environment \\ \hline
 & 275 & Environment \\ \hline
324 & 324 & Environment \\ \hline
\end{tabular}
\label{tab:config_1_abr}
\end{table}
Table \ref{tab:config_1_abr} lists all the modal frequencies that were identified in Configuration 1. The difference in the number of frequencies in the in-track and across-track dimensions suggests that the ambient environment in the cleanroom may have more seismic activity normal to the ground. In the across-track dimension, aside from those 35.02 Hz, and 120.10 Hz and their resonances at 70.46 Hz and 240.12 Hz, the rest of the observed frequencies appear to occur at irregular intervals. By subtracting out the modal frequencies present here from any tests run with the bearing off, we expect that the difference in the PSD plots represents only contributions from the satellite's vibrational sources.

\subsection{2 - Air Bearing off, Bus Electronics on, Reaction Wheels off, Cryocooler and Payload on}
Configuration 2 added the contribution from the cryocooler. The variance for the frame rate jitter was 0.0037 arcsec$^2$, which yields 1$\sigma$, 2$\sigma$, and 3$\sigma$ values of  0.06 arcsec, 0.12 arcsec, and 0.18 arcsec respectively. 

The identified modal frequencies found for this configuration are listed in Table \ref{tab:config_2_abr} where frequencies that were still present from the previous configuration but have different sources are highlighted in orange and new frequencies that were not present in the previous configuration are highlighted in green. In this configuration, they were determined to be contributions from the cryocooler by comparing the amplitude and FWHM values in Tables \ref{tab:config_1_freqs} and \ref{tab:config_2_freqs} in the Appendix. Because the cryocooler was operated at 60 Hz, the fundamental modal frequency as well as the resonance frequencies occurred in multiples of 60 in both the in-track and across-track dimensions. There was also a new modal frequency at 16 Hz in the in-track dimension as well as modal frequencies that are not multiples of 60 Hz, including 35 Hz, 93 Hz, and 460 Hz in the across-track dimension, and 340 Hz and 460 Hz in the in-track dimension.  
 
\begin{table}[hbt!]
\centering
\caption{Identified modal frequencies and their sources for Configuration 2. The vibrations from the cryocooler introduce new sets of frequencies in the system in both the in-track and across dimensions, especially at multiples of 60 Hz (the operating frequency of the cryocooler).}
\renewcommand*{\arraystretch}{1.1}
\begin{tabular}{|c|c|c|}
\hline
\begin{tabular}[c]{@{}c@{}}In-Track \\ Frequencies \\ Identified \\ {[}Hz{]}\end{tabular} & \begin{tabular}[c]{@{}c@{}}Across-Track \\ Frequencies\\ Identified \\ {[}Hz{]}\end{tabular} & Source \\ \hline
\cellcolor[HTML]{9AFF99}16.28 &  & Cryocooler \\ \hline
 & \cellcolor[HTML]{FFCE93}35.94 & \begin{tabular}[c]{@{}c@{}}Cryocooler,\\ Environment\end{tabular} \\ \hline
\cellcolor[HTML]{9AFF99}60.07 & \cellcolor[HTML]{9AFF99}60.06 & Cryocooler \\ \hline
 & \cellcolor[HTML]{9AFF99}93.02 & Cryocooler \\ \hline
\cellcolor[HTML]{9AFF99}120.11 & \cellcolor[HTML]{FFCE93}120.11 & \begin{tabular}[c]{@{}c@{}}Cryocooler, \\ Environment\end{tabular} \\ \hline
\cellcolor[HTML]{9AFF99}180.13 & \cellcolor[HTML]{FFCE93}180.13 & \begin{tabular}[c]{@{}c@{}}Cryocooler,\\ Environment\end{tabular} \\ \hline
\cellcolor[HTML]{FFCE93}240.13 & \cellcolor[HTML]{FFCE93}240.11 & \begin{tabular}[c]{@{}c@{}}Cryocooler,\\ Environment\end{tabular} \\ \hline
\cellcolor[HTML]{9AFF99}300.13 & \cellcolor[HTML]{9AFF99}300.13 & Cryocooler \\ \hline
\cellcolor[HTML]{9AFF99}340.19 &  & Cryocooler \\ \hline
\cellcolor[HTML]{9AFF99}360.10 & \cellcolor[HTML]{9AFF99}360.11 & Cryocooler \\ \hline
\cellcolor[HTML]{9AFF99}420.06 &  & Cryocooler \\ \hline
\cellcolor[HTML]{9AFF99}460.06 & \cellcolor[HTML]{9AFF99}460.08 & Cryocooler \\ \hline
\cellcolor[HTML]{9AFF99}479.99 & \cellcolor[HTML]{9AFF99}479.99 & Cryocooler \\ \hline
\end{tabular}
\label{tab:config_2_abr}
\end{table}

% As the bearing was not on for this configuration, we applied the same method to estimate the integration time jitter as discussed in the previous section. The resulting histogram is shown in Figure \ref{fig:config_2_jitter_hist} and shows a mode at 0.5 arcsec with most values lying between 0 and 1.0 arcsec. There is a non-zero region that lies between the jitter requirement at 2.89 arcsec and 3.0 arcsec.  
% \begin{figure}[hbt!] 
%     \label{fig:int_time_hist}
%     \centering
%     \includegraphics[width = 0.4\textwidth]{figures/int_time_hist.png}
%     \caption{Histogram of integration time jitter values taken from one of the trials for Configuration 2. While retaining the right-skewness from Figure \ref{fig:config_1_jitter_hist}, the mode occurs approximately at 0.5 arcsec. Note the non-zero region outside of the jitter requirement.}
%     \label{fig:config_2_jitter_hist}
% \end{figure} 

\subsection{3 - Air Bearing on, Bus Electronics on, Reaction Wheels off, Cryocooler and Payload off}
To inspect the dynamics of the air bearing itself, measurements were made with no active vibrational sources from HyTI turned on. Similar to Configuration 1, this configuration offers baseline values for comparison of vibratory modal frequencies. Due to oscillations caused by air bearing imbalance, the noise floor generated is high - approximately 10$^{-2}$ arcsec$^2$/Hz, as shown in Figure \ref{fig:PSD_config_3} in the Appendix. The resulting frame rate variance was 7.68 arcsec$^2$ which yields 1$\sigma$, 2$\sigma$, and 3$\sigma$ values of  2.77 arcsec, 5.54 arcsec, and 8.31 arcsec respectively. While these values are not relevant from the mission perspective as there are no active sources of jitter, we will compare them to the values from the next configuration to gain insight into the consistency of the metrology system. 

There was only one modal frequency identified by the algorithm in Table \ref{tab:config_3_abr}. However, this modal frequency is shown to be noise from sensor electronics for measurements recorded by the K-Cube when the laser drifts off of the sensor area of the detector. This peak occurs at 120 Hz because of the flicker from the fluorescent lighting in the cleanroom and is highlighted in orange to indicate that while it was present in Configuration 2, the source is not the same. 

\begin{table}[hbt!]
\centering
\caption{Identified modes and their sources for Configuration 3. Only a mode generated by electronic noise is present.}
\renewcommand*{\arraystretch}{1.1}
\begin{tabular}{|c|c|c|}
\hline
\begin{tabular}[c]{@{}c@{}}In-Track \\ Frequencies\\ Identified \\ {[}Hz{]}\end{tabular} & \begin{tabular}[c]{@{}c@{}}Across-Track \\ Frequencies\\ Identified \\ {[}Hz{]}\end{tabular} & Source \\ \hline
\cellcolor[HTML]{FFCE93}120.03 & \cellcolor[HTML]{FFCE93}120.03 & Noise \\ \hline
\end{tabular}
\label{tab:config_3_abr}
\end{table}

\subsection{4 - Air Bearing on, Bus Electronics on, ADCS Electronics on, Reaction Wheels off, Cryocooler and Payload off}
In this configuration, the ADCS electronics were turned on, but the reaction wheels, cryocooler, and payload remained inactive. The resulting frame rate variance was 1.85 arcsec$^2$ which yields 1$\sigma$, 2$\sigma$, and 3$\sigma$ values of 1.36 arcsec, 2.72 arcsec, and 4.08 arcsec respectively. These values were a slight decrease from the results of the previous configuration and are the result of the air bearing settling. This is also visible as a slight decrease in the baseline noise level in the PSD (Figure \ref{fig:PSD_config_4}) as this configuration was tested immediately after the previous one. The only modal frequency that was identified in this configuration shown in Table \ref{tab:config_4_abr} was the one generated by noise as in Configuration 3. Both have a comparable amplitude and FWHM as shown in Table \ref{tab:config_4_freqs} in the Appendix. 

\begin{table}[hbt!]
\centering
\caption{Identified modal frequencies and their sources for Configuration 4. There are no changes in the identified from Configuration 3.}
\renewcommand*{\arraystretch}{1.1}
\begin{tabular}{|c|c|c|}
\hline
\begin{tabular}[c]{@{}c@{}}In-Track \\ Frequencies\\ Identified \\ {[}Hz{]}\end{tabular} & \begin{tabular}[c]{@{}c@{}}Across-Track \\ Frequencies\\ Identified \\ {[}Hz{]}\end{tabular} & Source \\ \hline
\cellcolor[HTML]{FFFFFF}120.08 & \cellcolor[HTML]{FFFFFF}120.08 & Noise \\ \hline
\end{tabular}
\label{tab:config_4_abr}
\end{table}

%The tile rate jitter remains relatively unchanged from the last configuration with the variance being 20.38 arcsec$^2$, and the 1$\sigma$, 2$\sigma$, and 3$\sigma$ values being 4.51 arcsec, 9.02 arcsec, and 13.54 arcsec respectively. 

\subsection{5 - Air Bearing on, Bus Electronics on, ADCS Electronics on, x-direction Reaction Wheel spinning at 3000 rpm, Cryocooler and Payload off}
The air bearing remained on while the x-direction reaction wheel was spun at a constant 3000 rpm and the cryocooler and payload remained off. The variance for the frame rate jitter is 0.35 arcsec$^2$, which yields 1$\sigma$, 2$\sigma$, and 3$\sigma$ values of 0.60 arcsec, 1.19 arcsec, and 1.79 arcsec respectively. %For the tile rate, the variance decreases to 13.01 arcsec$^2$ with 1$\sigma$, 2$\sigma$, and 3$\sigma$ values of 3.61 arcsec, 7.21 arcsec, and 10.82 arcsec respectively. 
The 3$\sigma$ frame rate jitter value is smaller than the HyTI mission requirement for the integration time jitter.%We observe a decrease in the baseline values which is most significant in the across-track dimension through a comparison of Figures \ref{fig:PSD_config_5} and \ref{fig:PSD_config_4}. Whether this is due to the air bearing settling or the increase in internal angular momentum from the reaction wheel stabilizing the system has yet to be determined. Again, the only identified modal frequencies identified in Table \ref{tab:config_5_abr} are from the electronic noise.

\begin{table}[hbt!]
\centering
\caption{Identified modes and their sources for Configuration 5. There are no changes in the identified modes or their sources from Configuration 4.}
\renewcommand*{\arraystretch}{1.1}
\begin{tabular}{|c|c|c|}
\hline
\begin{tabular}[c]{@{}c@{}}In-Track \\ Frequencies\\ Identified \\ {[}Hz{]}\end{tabular} & \begin{tabular}[c]{@{}c@{}}Across-Track \\ Frequencies\\ Identified \\ {[}Hz{]}\end{tabular} & Source \\ \hline
\cellcolor[HTML]{FFFFFF}120.12 & \cellcolor[HTML]{FFFFFF}120.11 & Noise \\ \hline
\end{tabular}
\label{tab:config_5_abr}
\end{table}

The expected vibratory modal frequency in the in-track direction near 50 Hz that would be generated by the x-direction reaction wheel is not present and could be buried in the baseline nose.

\subsection{6 - Air Bearing on, Bus Electronics on, ADCS Electronics on, z-direction Reaction Wheel spinning at 3000 rpm, Cryocooler and Payload off}
The air bearing remained on, the x-direction reaction wheel was turned off and the only z-direction reaction wheel was spun at a constant 3000 rpm. The variance for the frame rate jitter is 0.021 arcsec$^2$, which yields 1$\sigma$, 2$\sigma$, and 3$\sigma$ values of 0.14 arcsec, 0.29 arcsec, 0.44 arcsec respectively, significantly less than previous trials and may suggest that the addition of angular momentum in the along the optical axis helps to stabilize the system more than the jitter it generates. The identified modal frequencies are listed in Table \ref{tab:config_6_abr}. There is no peak from the noise at 120 Hz in both the in-track and across-track dimensions. This is because the oscillations of the air bearing were constrained to the sensor area of the detector. From the PSD in Figure \ref{fig:PSD_config_6}, we observe a further decrease in the baseline values.

\begin{table}[hbt!]
\centering
\caption{Identified modal frequencies and their sources for Configuration 6. The mode at 120 Hz is not present as in this trial and we see a mode due to the low-frequency drift of the air bearing at 0.03 Hz and two frequencies, one at 11 Hz and the other at 58 Hz due to the activation of the z-direction reaction wheel.}
\renewcommand*{\arraystretch}{1.1}
\begin{tabular}{|c|c|c|}
\hline
\begin{tabular}[c]{@{}c@{}}In-Track \\ Frequencies\\ Identified \\ {[}Hz{]}\end{tabular} & \begin{tabular}[c]{@{}c@{}}Across-Track \\ Frequencies\\ Identified \\ {[}Hz{]}\end{tabular} & Source \\ \hline
\cellcolor[HTML]{9AFF99}0.03 & \cellcolor[HTML]{FFFFFF} & Air Bearing \\ \hline
 & \cellcolor[HTML]{9AFF99}11.12 & \begin{tabular}[c]{@{}c@{}}z-direction\\ Reaction Wheel\end{tabular} \\ \hline
 & \cellcolor[HTML]{9AFF99}58.27 & \begin{tabular}[c]{@{}c@{}}z-direction\\ Reaction Wheel\end{tabular} \\ \hline
\end{tabular}
\label{tab:config_6_abr}
\end{table}
\begin{figure*}[hbt!]
    \centering
    \includegraphics[width = 7 in]{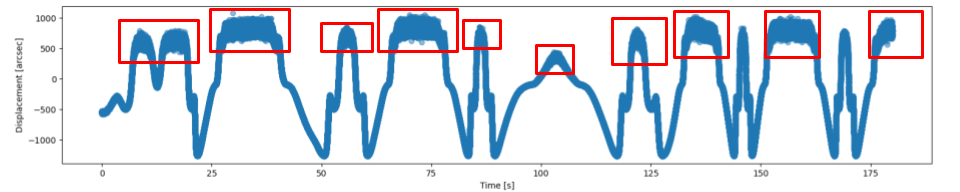}
    \caption{Data from one trial of this Configuration 3 in the in-track dimension demonstrating the variability of the data due to the air bearing imbalance. The noise seen at 120 Hz in both the in and across-track dimensions is generated when the laser spot drifts off of the position sensing detector and measurements are recorded. This occurs in the regions outlined with red boxes in the time series and manifests itself as a modal frequency at 120 Hz in the PSD.}
    \label{fig:bad_data}
\end{figure*}

In the in-track dimension, there was only one modal frequency identified at 0.03 Hz due to the low-frequency air-bearing drift. In the across-track dimension, there were two modal frequencies were identified, one at 11 Hz and another at 58 Hz. We attribute the modal frequency at 58 Hz to the spinning of the reaction wheel but would suggest that the wheel is not accurately spinning at 3000 rpm, which would be 50 Hz from the datasheet. Future trials hope to compare these results with rotor encoder data and compare the values of the reaction wheel angular rates.  %The variance for the tile rate jitter is 2.89 arcsec$^2$, which yields 1$\sigma$, 2$\sigma$, and 3$\sigma$ values of 1.70 arcsec, 3.40 arcsec, and 5.10 arcsec. Similar to Configuration 5, the 3$\sigma$ frame rate jitter was within the integration time jitter requirement for HyTI, as well as a three order-of-magnitude lower 3$\sigma$ value for the tile rate jitter.   

\subsection{7 - Air Bearing on, Bus Electronics on, ADCS Electronics on, x and z-direction Reaction Wheel spinning at 3000 rpm, Cryocooler and Payload off}
The air bearing remained on and the x-direction wheel was turned on to spin at the same time as the z-direction reaction wheel at a constant 3000 rpm while the cryocooler and payload remained off. The variance for the frame rate jitter is  0.019 arcsec$^2$, which yields 1$\sigma$, 2$\sigma$, and 3$\sigma$ values of 0.14 arcsec, 0.28 arcsec, 0.42 arcsec respectively - similar to Configuration 6. Identified modal frequencies are presented in Table \ref{tab:config_7_abr}.

\begin{table}[hbt!]
\centering
\caption{Identified modal frequencies and their sources for Configuration 7. Highlighted is the addition of two new modes.}
\renewcommand*{\arraystretch}{1.1}
\begin{tabular}{|c|c|c|}
\hline
\begin{tabular}[c]{@{}c@{}}In-Track \\ Frequency \\ Identified \\ {[}Hz{]}\end{tabular} & \begin{tabular}[c]{@{}c@{}}Across-Track \\ Frequency\\ Identified \\ {[}Hz{]}\end{tabular} & Source \\ \hline
\cellcolor[HTML]{FFFFFF}0.02 & \cellcolor[HTML]{FFFFFF} & Air Bearing \\ \hline
 & \cellcolor[HTML]{FFFFFF}10.40 & \begin{tabular}[c]{@{}c@{}}z-direction\\ Reaction \\ Wheel\end{tabular} \\ \hline
 & \cellcolor[HTML]{FFFFFF}58.52 & \begin{tabular}[c]{@{}c@{}}z-direction\\ Reaction \\ Wheel\end{tabular} \\ \hline
\cellcolor[HTML]{9AFF99}209.66 & \cellcolor[HTML]{9AFF99}209.63 & \begin{tabular}[c]{@{}c@{}}x- and z-direction\\ Reaction \\ Wheel\end{tabular} \\ \hline
\cellcolor[HTML]{9AFF99}368.35 &  & \begin{tabular}[c]{@{}c@{}}x- and z-direction\\ Reaction \\ Wheel\end{tabular} \\ \hline
\end{tabular}
\label{tab:config_7_abr}
\end{table}

In the in-track dimension, another modal frequency was identified at 0.02 Hz and can reasonably be considered the same modal frequency as one identified in Configuration 6 from the low-frequency drift of the air bearing. The modal frequencies at 10 Hz and 58 Hz remain present from the previous trial and we observed the addition of two higher frequency modal frequencies at 208 and 368 that were not present in Configuration 5 or Configuration 6. As these values are not multiples of 58 Hz (or 50 Hz), this demonstrates the benefits of investigating system dynamics with a fully integrated system.   %The variance for the tile rate jitter is 3.01 arcsec$^2$ which yields 1$\sigma$, 2$\sigma$, and 3$\sigma$ values of 1.74 arcsec, 3.47 arcsec, and 5.21 arcsec. The $3\sigma$ pointing jitter meets the HyTI requirement and tile rate jitter is remains three orders of magnitude underestimated.

The values of the frame rate jitter calculated for all configurations are summarized in Table \ref{tab:jitter_vals}. \textbf{The frame rate jitter values from configurations 5 - 7 are within the integration time jitter requirement for HyTI at the 3$\sigma$ level}.% The 3$\sigma$ tile rate jitter values from these configurations are also lower than those obtained from free-body simulations by approximately three orders of magnitude, even when dominated by noise.}
\begin{table}[hbt!]
\centering
\caption{Summary of calculated frame rate jitter values. All the 3$\sigma$ values are less than the integration time requirement of 2.89 arcsec.}
\renewcommand*{\arraystretch}{1.1}
\begin{tabular}{|c|c|c|c|}
\hline
Configuration & \begin{tabular}[c]{@{}c@{}}1$\sigma$\\ {[}arcsec{]}\end{tabular} & \begin{tabular}[c]{@{}c@{}}2$\sigma$\\ {[}arcsec{]}\end{tabular} & \begin{tabular}[c]{@{}c@{}}3$\sigma$\\ {[}arcsec{]}\end{tabular} \\ \hline
1 & 0.01 & 0.03 & 0.04 \\ \hline
2 & 0.06 & 0.12 & 0.18 \\ \hline
3 & 2.27 & 5.54 & 8.31 \\ \hline
4 & 1.36 & 2.72 & 4.08 \\ \hline
5 & 0.60 & 1.19 & 1.79 \\ \hline
6 & 0.14 & 0.29 & 0.44 \\ \hline
7 & 0.14 & 0.28 & 0.42 \\ \hline
\end{tabular}
\label{tab:jitter_vals}
\end{table}
%%%%%%%%%%%%%%%%%%%%%%%%%%%%%%%%%%%%%%%%
\section{Limitations} \label{sec:limitations}
%%%%%%%%%%%%%%%%%%%%%%%%%%%%%%%%%%%%%%%%
There are several limitations that we are devoting further work to address. At a campaign and logistical level, the fully integrated status of HyTI means that it is currently undergoing validation testing, and data collection for this endeavor was concurrent with the ADCS testing. We are hoping that the extension of HyTI's delivery date will present the opportunity to add to the configurations listed in Table \ref{tab:sources_tried} and allow us to analyze the jitter with the reaction wheels and cryocooler active. 

At an experimental level, imbalances with the air bearing as mentioned in Section \ref{sec:testbed} resulted in noisy data. Better characterizing the oscillations caused by imbalances or eliminating them would yield cleaner results. While these can be filtered out after the fact, the primary issue with these oscillations is that they lead they carry the laser beam off of the sensor resulting in PSD's like those shown in Figure \ref{fig:bad_data}. Cleaner results will allow us to apply the technique from configurations 1 and 2 to estimate the integration time jitter for configurations using the air bearing. Because this results in a slow drift of the air bearing, this noise dominates the lower frequencies (less than about 1 Hz). 

Secondly, the use of the support jig adds approximately 3 kg of mass and increases the effective moment of inertia of the entire system. The increased mass in the y-direction in HyTI's coordinate frame changes the frequency response of the system. Resulting in the shifting of modal frequencies and their amplitudes and altering principle axes of the entire system. This leads to a systematic underestimation of jitter and work to quantify this underestimation will serve to better understand the response scaling so that results are more representative of HyTI. 

At the metrology system level, the main limitation is the ADC sampling frequency. Since the integration time jitter will be observable at 2 kHz, direct measurements require an ADC that can sample at least at a Nyquist rate of 4 kHz over three channels - requiring a device that samples at least 12 kHz. The lateral effect position sensor provides an upper limit on the required ADC of 15 kHz.

%%%%%%%%%%%%%%%%%%
\section{Conclusions} \label{sec:conclusions}
%%%%%%%%%%%%%%%%%%
We present the optical metrology setup and methodology behind a novel approach to investigate system dynamics and conduct jitter analysis in the frequency domain. We use these to explore the jitter characteristics of HyTI due to its reaction wheel control system and cryocooler and determine if it satisfies mission requirements. We also identify modal frequencies associated with different vibratory sources. This method is a proposed low-cost and precise way of characterizing jitter in cubesats (and smallsats) that does not require force transducers nor a finite element model. It can be adapted to study jitter at a component or sub-assembly level if desired and scalable to measure in three degrees of freedom. 

Future work will be aimed at addressing some of the limitations mentioned previously. Given the data that was collected, more investigation is necessary to characterize the decrease in the baseline levels between Configurations 3 and 7. This could be the air bearing settling as airflow becomes less turbulent or an increase in the internal angular momentum of the satellite as alluded to in Configuration 5. New configurations will be tested, allowing for more information on the cryocooler's contribution to the jitter values and system dynamics. We also wish to better understand the dynamics of different reaction wheel speeds and understand why a modal frequency at 58 Hz is present instead of 50 Hz. Better characterizing the damping properties of the clamp will allow us to adjust the results presented here to be more representative of the actual satellite's response. Ultimately, we wish to compare these results with in-orbit data from HyTI after its launch to better inform the development of this metrology system.  

%%%%%%%%%%%%%%%%%%%%%%%%%%%%%%%%%%%%%%%%%%%%%%%%%%%%%%%%%%%%%%%%%%%%%%%%%%%%%%%%%%%%%%%%%%%%%%%%%%%%%%
\bibliographystyle{IEEEtran}
\bibliography {refs}


% \bibitem{ITAR}
% U.S. Munitions List, Sections 38 and 47(7) of the Arms Export Control Act (22 U.S.C 2778 and 2794(7).

% \bibitem{AeroConf}
% Aerospace Conference Web site: \underline{www.aeroconf.org}.

% \end{thebibliography}

% %%%%%%%%%%%%%%%%%%%%%%%%%%%%%%%%%%%%%%%%%%%%%%%%%%%%%%%%%%%%%%%%%%%%%%%%%%%%%%%%%%%%%%%%%%%%%%%%%%%%%%
\thebiography
%% This biostyle allows you to insert your photo size 1in X 1.25in
\begin{biographywithpic}
{Chase Urasaki}{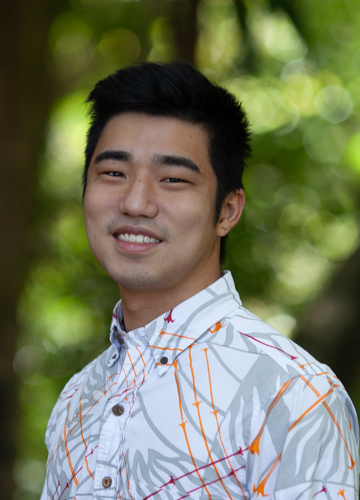}
received his B.S. degrees in Astrophysics and Mathematics from the University of Hawai\okina i at M\=anoa in 2019. He is currently pursuing an M.S. in Electrical Engineering at the University of Hawai\okina i at M\=anoa and a Graduate Research Assistant with the Hawai\okina i Space Flight Lab.
\end{biographywithpic} 

\begin{biographywithpic}
{Frances Zhu}{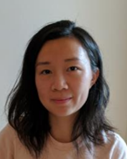}
earned her B.S. in Mechanical and Aerospace Engineering from Cornell University, Ithaca in 2014 and a Ph.D. in Aerospace Engineering at Cornell in 2019. Dr. Zhu was a NASA Space Technology Research Fellow. Since 2020, she has been an assistant research professor with the University of Hawai\okina i, specializing in machine learning, dynamics, systems, and controls engineering. She is also the deputy director of the Hawaii Space Grant Consortium and graduate cooperating faculty with the following departments: mechanical engineering, electrical engineering, information and computer science, and earth science.

\end{biographywithpic}

\begin{biographywithpic}
    {Michael Bottom}{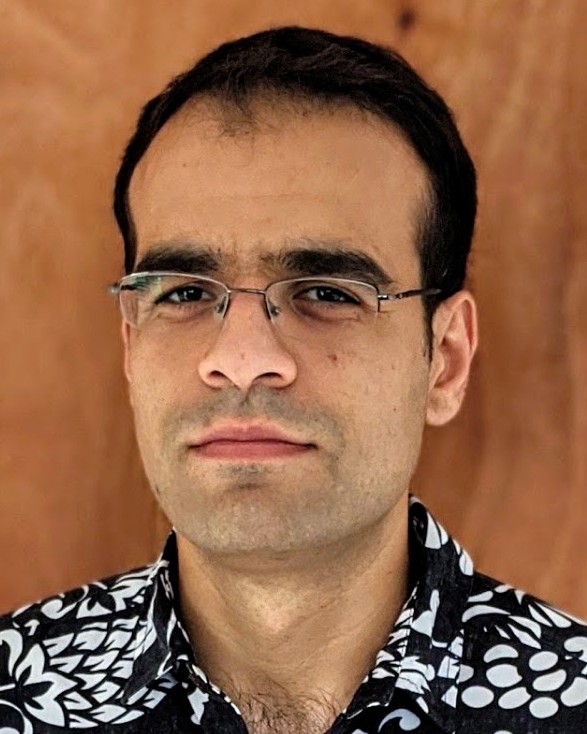}
     received his B.A. degree in math and physics from Columbia University in 2008 and his Ph.D. in Astrophysics from the California Institute of Technology in 2016. He was an optical engineer at NASA’s Jet Propulsion Lab from 2016-2019 and is currently an assistant professor of astronomy at the University of Hawai\okina i at M\=anoa. His research interests are mainly in the field of exoplanet imaging and spectroscopy, including infrared detector development, wavefront sensing, and space instrumentation.
\end{biographywithpic}

\begin{biographywithpic}
    {Miguel Nunes}{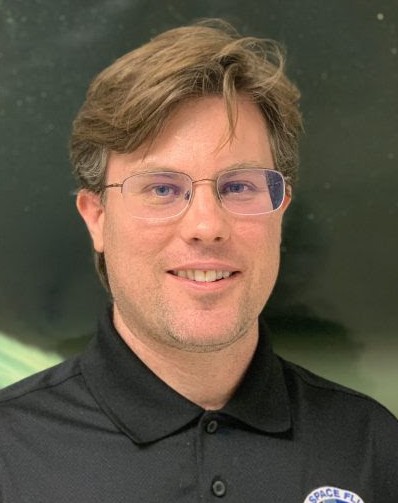}
    is an Assistant Researcher at the Institute of Geophysics and Planetology (HIGP) and Deputy Director of the Hawaii Space Flight Laboratory (HSFL). He received his Aerospace Engineering Degree from the Instituto Superior Técnico, Portugal, and his Ph.D. at the University of Hawai\okina i at M\=anoa in 2016. During his Ph.D. research, Miguel developed a Multi-Agent Robotic System and GNC algorithms for distributed space missions. He worked as the Systems Engineer and Assistant Project Manager for the Neutron-1 mission. His current work is focused on small satellite mission development with applications to space science instrumentation, such as the NASA HyTI mission, where he serves as the Systems Engineer and Deputy PI. Miguel is also a co-inventor of the COSMOS software framework (https://github.com/hsfl/cosmos/) an open-source flight software and toolchain for satellite operations and control. 
\end{biographywithpic}

\begin{biographywithpic}
    {Aidan Walk}{figures/aw.jpg}
    is a Telescope Operator at Subaru Telescope, NAOJ. He obtained his B.S. in Astronomy from the University of Hawaii at Hilo in 2022, where he assisted with research involving astrometric calibration, telescope operation, and IR Detector development.
\end{biographywithpic}

%%%%%%%%%%%%%%%%%%%%%%%%%%%%%%%%%%%%%%%%%%%%%%%%%%%%%%%%%%%%%%%%%%%%%%%%%%%%%%%%%%%%%%%%%%%%%%%%%%%%%%
\acknowledgments
The authors would like to acknowledge funding and support from the Hawai\okina i Space Grant Consortium, the Hawai\okina i Space Flight Lab, and HyTI's Principal Investigator, Dr. R. Wright for which this work is possible.

\newpage

\onecolumn
\appendix{}
% note there is no {} to put a title. Each appendix has its own title
PSD plots and full tables for identified modal frequencies for each of the configurations listed in Table \ref{tab:sources_tried}. The across-track PSD is plotted in orange and the in-track PSDs are plotted in blue. The HyTI tile and frame rates are indicated by vertical dashed lines and error bars are omitted for clarity. Tables are formatted similarly to the text: A green highlight indicating a new modal frequency that was not there in the previous configuration, an orange highlight indicating a modal frequency that was in the previous configuration but was attributed to a different source, or no highlight where the modal frequency and source are the same as the previous configuration.  
\begin{figure*}[hbt!]
    \centering
    \includegraphics[width = 7in]{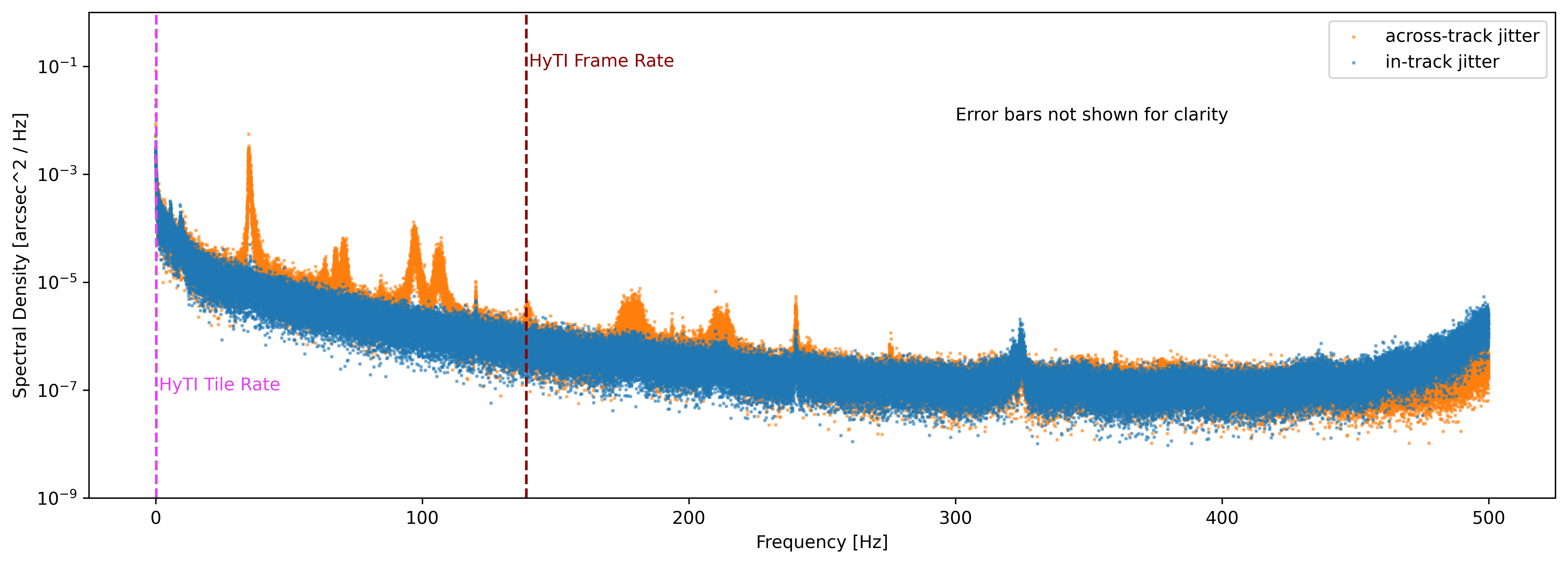}
    \caption{PSD from Configuration 1}
    \label{fig:PSD_config_1}
\end{figure*} 

\begin{table*}[hbt!] 
\centering
\caption{Identified modeal frequencies and their sources for configuration 1.}
\renewcommand*{\arraystretch}{1.3}
\begin{tabular}{|c|c|c|c|c|c|c|}
\hline
\begin{tabular}[c]{@{}c@{}}In-Track Frequencies \\ Identified {[}Hz{]}\end{tabular} &
  \begin{tabular}[c]{@{}c@{}}Amplitude\\ {[}arcsec$^2$/Hz{]}\end{tabular} &
  FWHM {[}Hz{]} &
  \begin{tabular}[c]{@{}c@{}}Across-Track Frequencies\\ Identified {[}Hz{]}\end{tabular} &
  \begin{tabular}[c]{@{}c@{}}Amplitude\\ {[}arcsec$^2$/Hz{]}\end{tabular} &
  FWHM {[}Hz{]} &
  Source \\ \hline
    &          &          & 34.94  & 1.75E-03 & 3.72E-01 & Environment \\ \hline
    &          &          & 70.46  & 2.25E-05 & 1.11E+00 & Environment \\ \hline
    &          &          & 97.15  & 5.70E-05 & 1.05E+00 & Environment \\ \hline
    &          &          & 106.25 & 2.13E-05 & 1.90E+00 & Environment \\ \hline
    &          &          & 120.10 & 4.60E-06 & 1.21E-01 & Environment \\ \hline
    &          &          & 139.56 & 1.15E-06 & 9.54E-01 & Environment \\ \hline
    &          &          & 179.36 & 2.91E-06 & 5.68E+00 & Environment \\ \hline
    &          &          & 211.88 & 1.27E-06 & 4.90E+00 & Environment \\ \hline
240.13 & 3.79E-07 & 2.26E-01 & 240.12 & 2.33E-06 & 2.41E-01 & Environment \\ \hline
    &          &          & 275.71 & 2.04E-07 & 5.19E-01 & Environment \\ \hline
324.54 & 5.36E-07 & 7.77E-01 & 324.58 & 2.27E-07 & 7.51E-01 & Environment \\ \hline
\end{tabular}
\label{tab:config_1_freqs}
\end{table*}
\newpage
\begin{figure*}[hbt!]
    \centering
    \includegraphics[width = 7in]{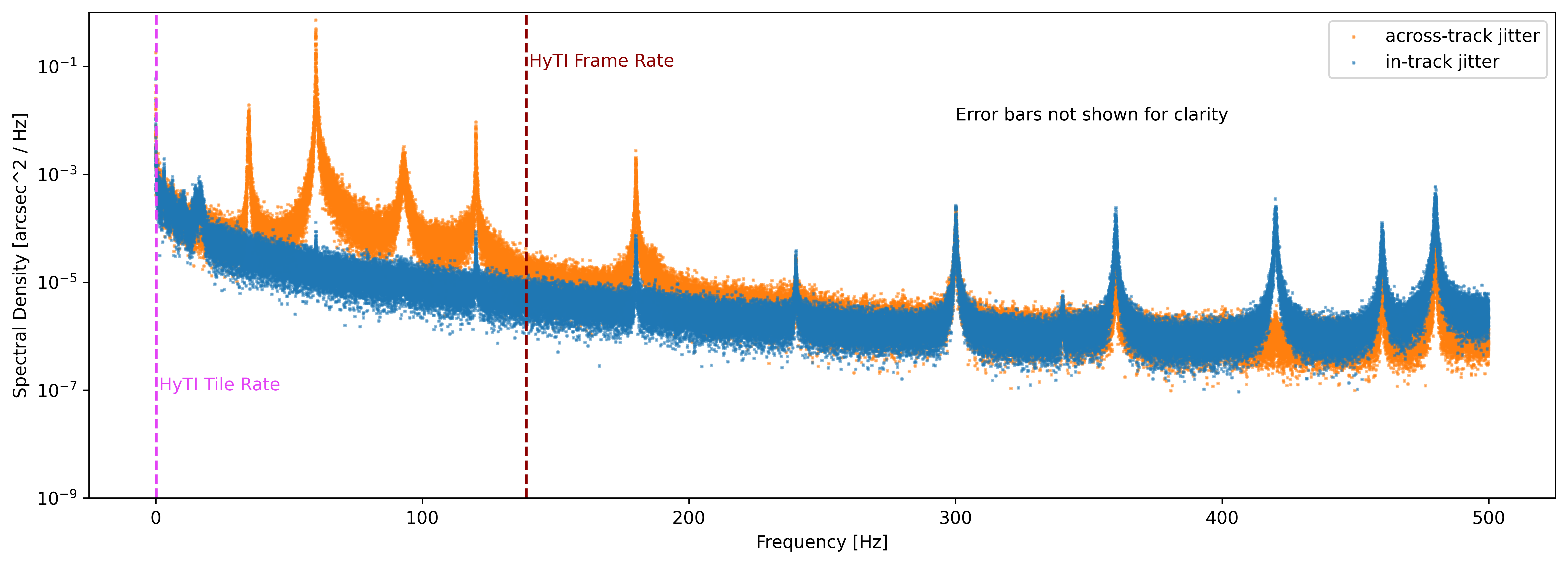}
    \caption{PSD from Configuration 2}
    \label{fig:PSD_config_2}
\end{figure*}


\begin{table*}[hbt!]
\centering
\caption{Identified modal frequencies and their sources for Configuration 2. Green highlights represent frequencies not present in the previous trial and orange highlights are frequencies that were present in the previous trial but differ significantly through a comparison of amplitudes.}
\renewcommand*{\arraystretch}{1.3}
\begin{tabular}{|c|c|c|c|c|c|c|}
\hline
\begin{tabular}[c]{@{}c@{}}In-Track Modes \\ Identified {[}Hz{]}\end{tabular} &
  \begin{tabular}[c]{@{}c@{}}Amplitude\\ {[}arcsec$^2$/Hz{]}\end{tabular} &
  FWHM {[}Hz{]} &
  \begin{tabular}[c]{@{}c@{}}Across-Track Modes\\ Identified {[}Hz{]}\end{tabular} &
  \begin{tabular}[c]{@{}c@{}}Amplitude\\ {[}arcsec$^2$/Hz{]}\end{tabular} &
  FWHM {[}Hz{]} &
  Source \\ \hline
\cellcolor[HTML]{9AFF99}16.28 &
  2.45E-04 &
  1.46E+00 &
   &
   &
   &
  Cryocooler \\ \hline
 &
   &
   &
  \cellcolor[HTML]{FFCE93}35.94 &
  1.02E-02 &
  2.57E-01 &
  \begin{tabular}[c]{@{}c@{}}Cryocooler,\\ Environment\end{tabular} \\ \hline
\cellcolor[HTML]{9AFF99}60.07 &
  3.73E-05 &
  9.41E-02 &
  \cellcolor[HTML]{9AFF99}60.06 &
  3.66E-01 &
  7.67E-02 &
  Cryocooler \\ \hline
 &
   &
   &
  \cellcolor[HTML]{9AFF99}93.02 &
  1.31E-03 &
  8.88E-01 &
  Cryocooler \\ \hline
\cellcolor[HTML]{9AFF99}120.11 &
  4.05E-05 &
  1.23E-01 &
  \cellcolor[HTML]{FFCE93}120.11 &
  4.94E-03 &
  1.14E-01 &
  \begin{tabular}[c]{@{}c@{}}Cryocooler, \\ Environment\end{tabular} \\ \hline
\cellcolor[HTML]{9AFF99}180.13 &
  3.98E-05 &
  1.65E-01 &
  \cellcolor[HTML]{FFCE93}180.13 &
  1.39E-03 &
  1.69E-01 &
  \begin{tabular}[c]{@{}c@{}}Cryocooler,\\ Environment\end{tabular} \\ \hline
\cellcolor[HTML]{FFCE93}240.13 &
  1.64E-05 &
  2.35E-01 &
  \cellcolor[HTML]{FFCE93}240.11 &
  1.36E-05 &
  2.34E-01 &
  \begin{tabular}[c]{@{}c@{}}Cryocooler,\\ Environment\end{tabular} \\ \hline
\cellcolor[HTML]{9AFF99}300.13 &
  1.31E-04 &
  2.80E-01 &
  \cellcolor[HTML]{9AFF99}300.13 &
  1.18E-04 &
  2.77E-01 &
  Cryocooler \\ \hline
\cellcolor[HTML]{9AFF99}340.19 &
  2.10E-06 &
  3.28E-01 &
   &
   &
   &
  Cryocooler \\ \hline
\cellcolor[HTML]{9AFF99}360.10 &
  1.14E-04 &
  3.54E-01 &
  \cellcolor[HTML]{9AFF99}360.11 &
  1.61E-05 &
  3.67E-01 &
  Cryocooler \\ \hline
\cellcolor[HTML]{9AFF99}420.06 &
  1.51E-04 &
  3.85E-01 &
   &
   &
   &
  Cryocooler \\ \hline
\cellcolor[HTML]{9AFF99}460.06 &
  6.20E-05 &
  4.49E-01 &
  \cellcolor[HTML]{9AFF99}460.08 &
  5.51E-06 &
  4.38E-01 &
  Cryocooler \\ \hline
\cellcolor[HTML]{9AFF99}479.99 &
  2.84E-04 &
  4.23E-01 &
  \cellcolor[HTML]{9AFF99}479.99 &
  3.10E-05 &
  4.20E-01 &
  Cryocooler \\ \hline
\end{tabular}
\label{tab:config_2_freqs}
\end{table*}

\newpage
\begin{figure*}[hbt!]
    \centering
    \includegraphics[width = 7in]{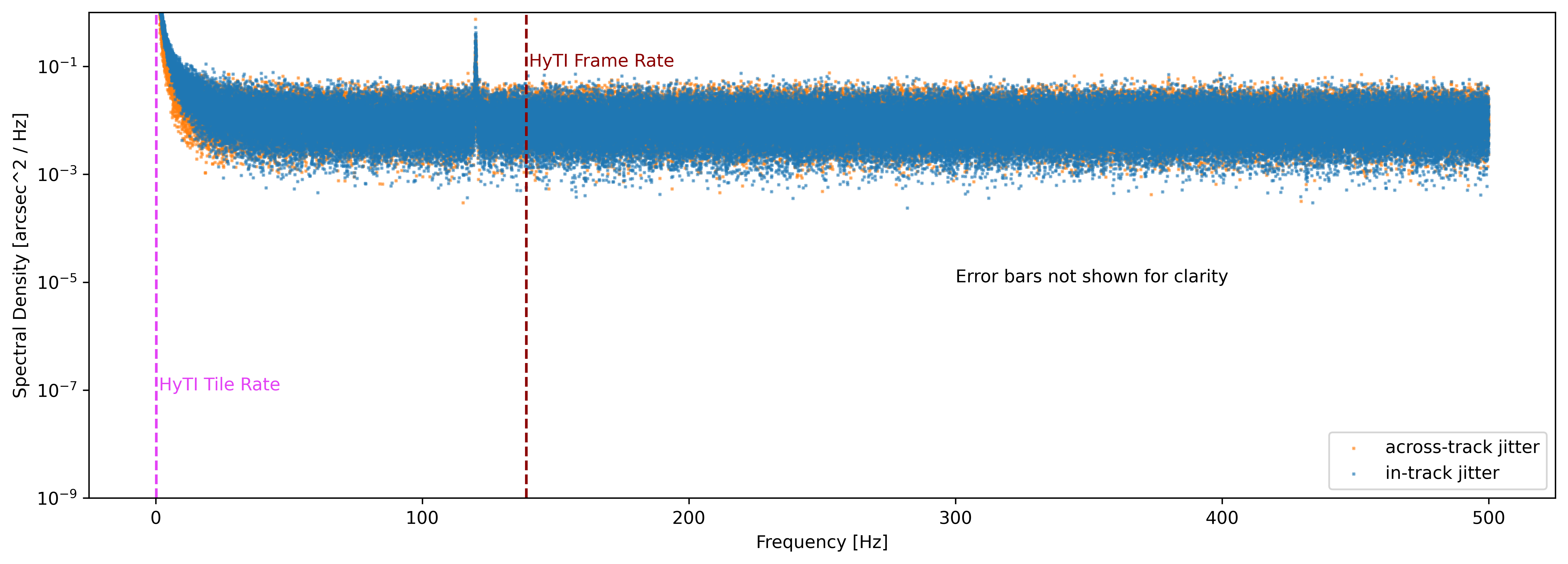}
    \caption{PSD from Configuration 3}
    \label{fig:PSD_config_3}
\end{figure*}

\begin{table*}[hbt!]
\centering
\caption{Identified modal frequencies and their sources for Configuration 3. Green highlights represent frequencies not present in the previous trial and orange highlights are frequencies that were present in the previous trial but differ significantly through a comparison of amplitudes.}
\renewcommand*{\arraystretch}{1.3}
\begin{tabular}{|c|c|c|c|c|c|c|}
\hline
\begin{tabular}[c]{@{}c@{}}In-Track Frequencies\\ Identified {[}Hz{]}\end{tabular} &
  \begin{tabular}[c]{@{}c@{}}Amplitude \\ {[}arcsec$^2$/Hz{]}\end{tabular} &
  FWHM {[}Hz{]} &
  \begin{tabular}[c]{@{}c@{}}Across-Track Frequencies\\ Identified {[}Hz{]}\end{tabular} &
  \begin{tabular}[c]{@{}c@{}}Amplitude \\ {[}arcsec$^2$/Hz{]}\end{tabular} &
  FWHM {[}Hz{]} &
  Source \\ \hline
\cellcolor[HTML]{FFCE93}120.03 &
  2.57E-01 &
  1.27E-1 &
  \cellcolor[HTML]{FFCE93}120.03 &
  2.00E-01 &
  1.36E-01 &
  Noise \\ \hline
\end{tabular}
\label{tab:config_3_freqs}
\end{table*}

\begin{figure*}[hbt!]
    \centering
    \includegraphics[width = 7in]{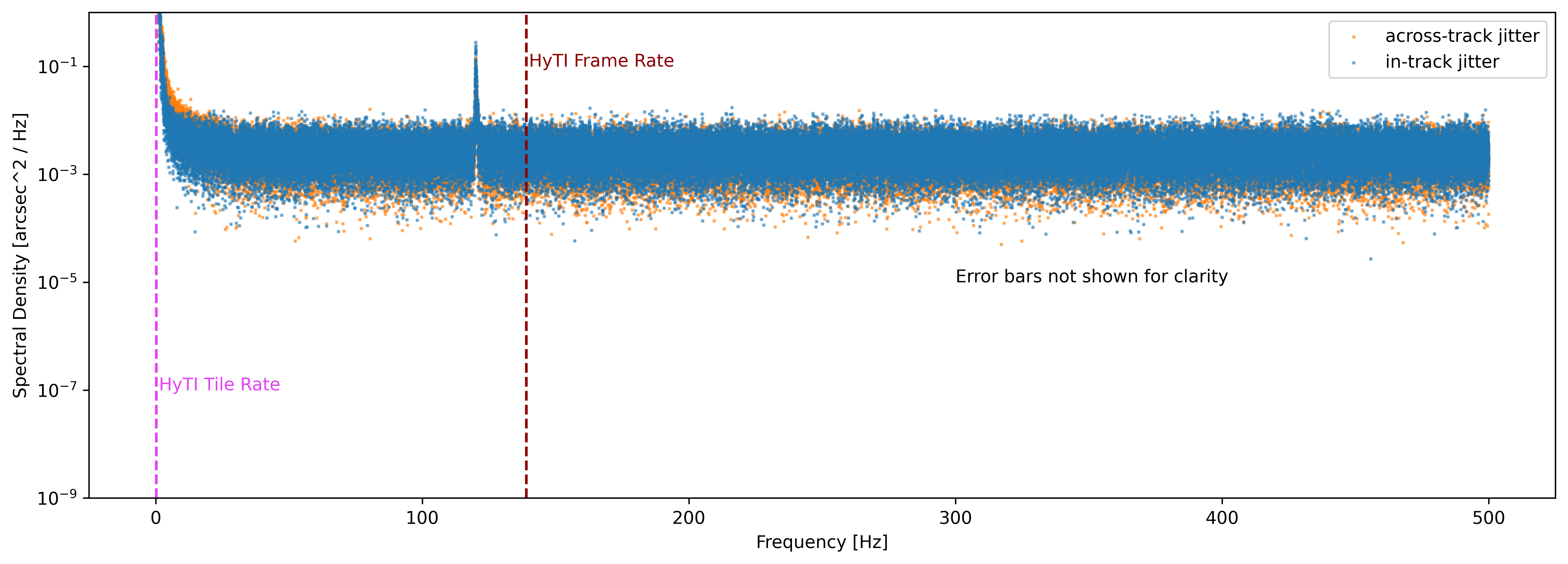}
    \caption{PSD from Configuration 4}
    \label{fig:PSD_config_4}
\end{figure*}

\begin{table*}[hbt!]
\centering
\caption{Identified modes and their sources for Configuration 4. No highlighted frequencies indicated that no frequency or sources have changed since the previous configuration.}
\renewcommand*{\arraystretch}{1.3}
\begin{tabular}{|c|c|c|c|c|c|c|}
\hline
\begin{tabular}[c]{@{}c@{}}In-Track Frequencies \\ Identified {[}Hz{]}\end{tabular} &
  \begin{tabular}[c]{@{}c@{}}Amplitude\\ {[}arcsec$^2$/Hz{]}\end{tabular} &
  FWHM {[}Hz{]} &
  \begin{tabular}[c]{@{}c@{}}Across-Track Frequencies\\ Identified {[}Hz{]}\end{tabular} &
  \begin{tabular}[c]{@{}c@{}}Amplitude\\ {[}arcsec$^2$/Hz{]}\end{tabular} &
  FWHM {[}Hz{]} &
  Source \\ \hline
\cellcolor[HTML]{FFFFFF}120.08 &
  1.35E-01 &
  1.10E-01 &
  \cellcolor[HTML]{FFFFFF}120.08 &
  8.83E-02 &
  1.17E-01 &
  Noise \\ \hline
\end{tabular}
\label{tab:config_4_freqs}
\end{table*}

\begin{figure*} [hbt!]
    \centering
    \includegraphics[width = 7in]{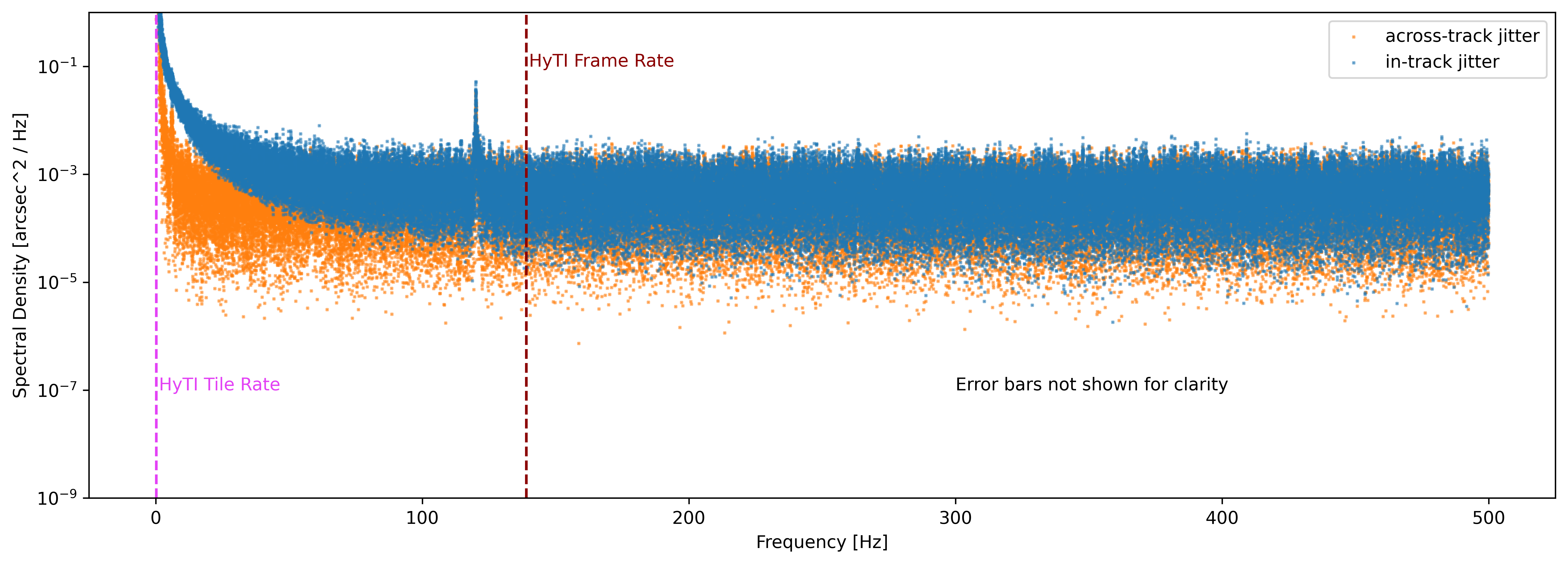}
    \caption{PSD from Configuration 5}
    \label{fig:PSD_config_5}
\end{figure*}

\begin{table*}[hbt!]
\centering
\caption{Identified modal frequencies and their sources for Configuration 5. No differences  were identified despite the addition of the x-direction reaction wheel spinning at 3000 rpm.}
\renewcommand*{\arraystretch}{1.3}
\begin{tabular}{|c|c|c|c|c|c|c|}
\hline
\begin{tabular}[c]{@{}c@{}}In-Track Frequencies\\ Identified {[}Hz{]}\end{tabular} &
  \begin{tabular}[c]{@{}c@{}}Amplitude\\ {[}arcsec$^2$/Hz{]}\end{tabular} &
  \begin{tabular}[c]{@{}c@{}}FWHM \\ {[}Hz{]}\end{tabular} &
  \begin{tabular}[c]{@{}c@{}}Across-Track Frequencies\\ Identified {[}Hz{]}\end{tabular} &
  \begin{tabular}[c]{@{}c@{}}Amplitude\\ {[}arcsec$^2$/Hz{]}\end{tabular} &
  \begin{tabular}[c]{@{}c@{}}FWHM \\ {[}Hz{]}\end{tabular} &
  Source \\ \hline
\cellcolor[HTML]{FFFFFF}120.12 &
  2.20E-02 &
  1.28E-01 &
  \cellcolor[HTML]{FFFFFF}120.11 &
  1.39E-02 &
  1.34E-01 &
  Noise \\ \hline
\end{tabular}
\label{tab:config_5_freqs}
\end{table*}

\begin{figure*}[hbt!]
    \centering
    \includegraphics[width = 7in]{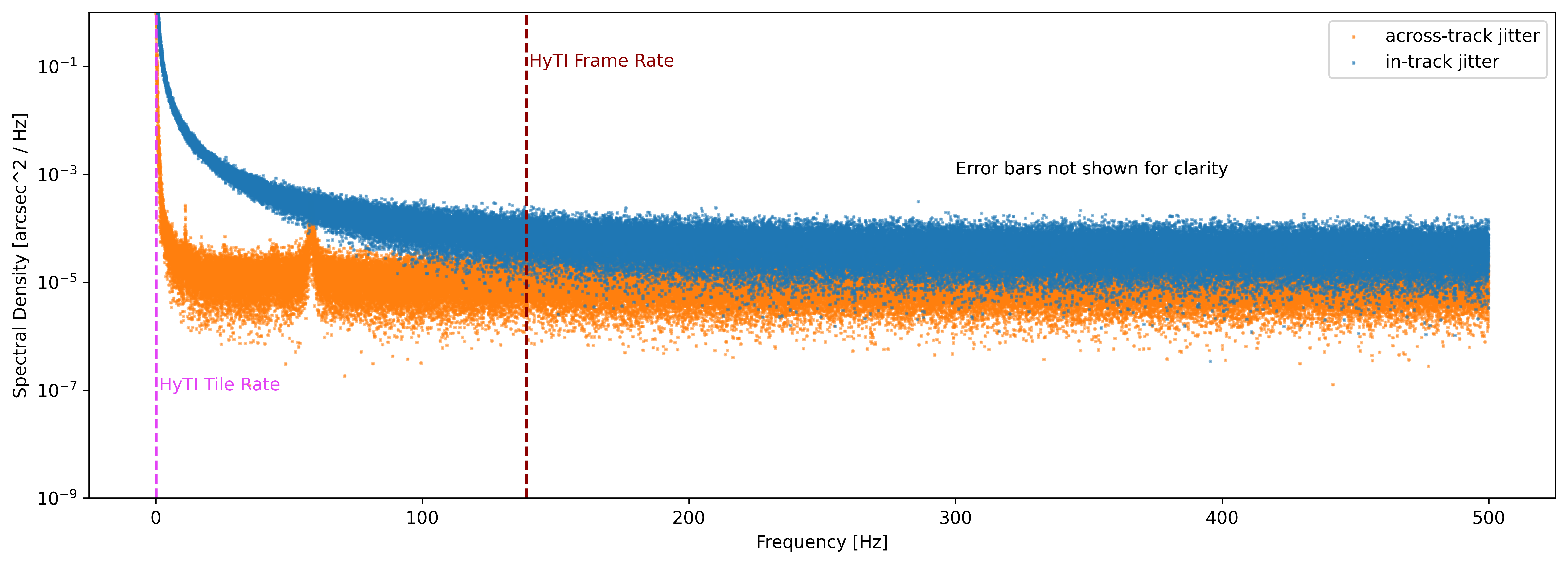}
    \caption{PSD from Configuration 6}
    \label{fig:PSD_config_6}
\end{figure*}

\begin{table*}[hbt!]
\centering
\caption{Identified modal frequencies and their sources for Configuration 6. The peak in the in-track dimension is due to the low-frequency oscillation of the air bearing. Both modes that are identified in the across-track dimension are from the z-direction reaction wheel.}
\renewcommand*{\arraystretch}{1.3}
\begin{tabular}{|c|c|c|c|c|c|c|}
\hline
\begin{tabular}[c]{@{}c@{}}In-Track Frequencies\\ Identified {[}Hz{]}\end{tabular} &
  \begin{tabular}[c]{@{}c@{}}Amplitude\\ {[}arcsec$^2$/Hz{]}\end{tabular} &
  \begin{tabular}[c]{@{}c@{}}FWHM\\ {[}Hz{]}\end{tabular} &
  \begin{tabular}[c]{@{}c@{}}Across-Track Frequencies\\ Identified {[}Hz{]}\end{tabular} &
  \begin{tabular}[c]{@{}c@{}}Amplitude\\ {[}arcsec$^2$/Hz{]}\end{tabular} &
  \begin{tabular}[c]{@{}c@{}}FWHM\\ {[}Hz{]}\end{tabular} &
  Source \\ \hline
\cellcolor[HTML]{9AFF99}0.03 &
  1.26E+04 &
  8.87E-05 &
  \cellcolor[HTML]{FFFFFF} &
   &
   &
  Air Bearing \\ \hline
 &
   &
   &
  \cellcolor[HTML]{9AFF99}11.12 &
  \multicolumn{1}{r|}{8.87E-05} &
  \multicolumn{1}{r|}{7.37E-02} &
  \begin{tabular}[c]{@{}c@{}}z-direction\\ Reaction Wheel\end{tabular} \\ \hline
 &
   &
   &
  \cellcolor[HTML]{9AFF99}58.27 &
  \multicolumn{1}{r|}{1.21E-04} &
  \multicolumn{1}{r|}{8.35E-01} &
  \begin{tabular}[c]{@{}c@{}}z-direction\\ Reaction Wheel\end{tabular} \\ \hline
\end{tabular}
\label{tab:config_6_freqs}
\end{table*}

\begin{figure*} [hbt!]
    \centering
    \includegraphics[width = 7in]{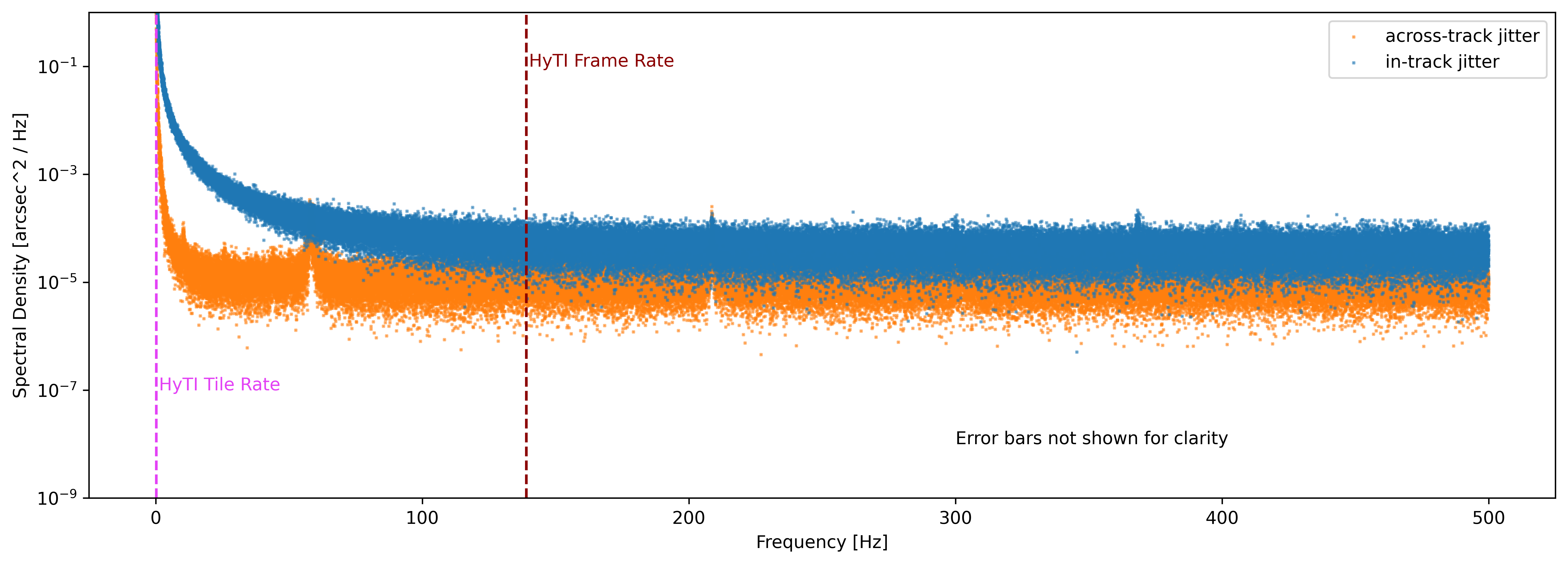}
    \caption{PSD from Configuration 7}
    \label{fig:PSD_config_7}
\end{figure*}
\begin{table*}[hbt!]
\centering
\caption{Identified modal frequencies and their sources for Configuration 7. Note the addition of frequencies that were not present in the previous trials.}
\renewcommand*{\arraystretch}{1.3}
\begin{tabular}{|c|c|c|c|c|c|c|}
\hline
\begin{tabular}[c]{@{}c@{}}In-Track Frequency \\ Identified {[}Hz{]}\end{tabular} &
  \begin{tabular}[c]{@{}c@{}}Amplitude\\ {[}arcsec$^2$/Hz{]}\end{tabular} &
  \begin{tabular}[c]{@{}c@{}}FWHM \\ {[}Hz{]}\end{tabular} &
  \begin{tabular}[c]{@{}c@{}}Across-Track Frequency\\ Identified {[}Hz{]}\end{tabular} &
  \begin{tabular}[c]{@{}c@{}}Amplitude\\ {[}arcsec$^2$/Hz{]}\end{tabular} &
  \begin{tabular}[c]{@{}c@{}}FWHM \\ {[}Hz{]}\end{tabular} &
  Source \\ \hline
\cellcolor[HTML]{FFFFFF}0.02 &
  \cellcolor[HTML]{FFFFFF}1.19E+04 &
  \cellcolor[HTML]{FFFFFF}4.79E-02 &
  \cellcolor[HTML]{FFFFFF} &
   &
   &
  Air Bearing \\ \hline
 &
   &
   &
  \cellcolor[HTML]{FFFFFF}10.40 &
  2.93E-05 &
  1.36E-01 &
  \begin{tabular}[c]{@{}c@{}}z-direction\\ Reaction Wheel\end{tabular} \\ \hline
 &
   &
   &
  \cellcolor[HTML]{FFFFFF}58.52 &
  9.89E-05 &
  9.19E-01 &
  \begin{tabular}[c]{@{}c@{}}z-direction\\ Reaction Wheel\end{tabular} \\ \hline
\cellcolor[HTML]{9AFF99}209.66 &
  3.56E-05 &
  3.18E-01 &
  \cellcolor[HTML]{9AFF99}209.63 &
  3.56E-05 &
  3.18E-01 &
  \begin{tabular}[c]{@{}c@{}}x- and z-direction\\ Reaction Wheel\end{tabular} \\ \hline
\cellcolor[HTML]{9AFF99}368.35 &
  5.10E-05 &
  5.04E-01 &
   &
   &
   &
  \begin{tabular}[c]{@{}c@{}}x- and z-direction\\ Reaction Wheel\end{tabular} \\ \hline
\end{tabular}
\label{tab:config_7_freqs}
\end{table*}

%%%%%%%%%%%%%%%%%%%%%%%%%%%%%%%%%%%%%%%%%%%%%%%%%%%%%%%%%%%%%%%%%%%%%%%%%%%%%%%%%%%%%%%%%%%%%%%%%
% % For a single appendix, use the \appendix{} keyword and do not use the \section command.

% \appendix{More Information}        % first appendix
% %%%%%%%%%%%%%%%%%%%%%%%%%%
% This is the first appendix. 

% \subsection{Comments}
% If you have only one appendix, use the ``appendix'' keyword.

% \subsection{More Comments}
% Use section and subsection keywords as usual.

% \section{Yet More Information}    % second appendix
% %%%%%%%%%%%%%%%%%%%%%%%%%%%%%%
% This is the second appendix.

\end{document}